\documentstyle[12pt]{article}

\title{\vspace*{-10mm}\hrulefill\\{\bf The Origin of the RNA World: 
a Kinetic Model}}
\author{ {\bf Jonathan AD Wattis}$^{\dag*}$ {\small and} 
{\bf Peter V Coveney}$^{\ddag*}$ \\
{\footnotesize $^\dag$Department of Theoretical Mechanics, University of 
Nottingham, University Park, Nottingham, NG7 2RD, U.K.} \\[-0.2ex] 
{\footnotesize $^\ddag$Schlumberger Cambridge Research, High Cross,
Madingley Road, Cambridge, CB3 0EL, U.K.,} \\[-0.8ex]
{\footnotesize and Department of Theoretical Physics, University of Oxford,
Keble Road, Oxford OX1 3NP, U.K., now at} \\[-0.8ex] 
{\footnotesize \hspace*{-6mm}Centre for Computational Science, Queen Mary and Westfield
College, University of London, Mile End Road, London, E1 4NS} \\[-0.2ex] 
{\footnotesize $^\dag$\verb$Jonathan.Wattis@nottingham.ac.uk$ 
\hspace{8mm} $^\ddag$\verb$coveney@cambridge.scr.slb.com$}\\[-1ex]}
\date{{\small\sl 6$^{{\rm th}}$ January, 1999.}\\ \hrulefill}

\newcommand{\beqan}{\begin{eqnarray*}}
\newcommand{\eeqan}{\end{eqnarray*}}
\newcommand{\beqa}{\begin{eqnarray}}
\newcommand{\eeqa}{\end{eqnarray}}
\newcommand{\beq}{\begin{equation}}
\newcommand{\eeq}{\end{equation}}
\newcommand{\rec}[1]{\mbox{$\frac{1}{#1}$}}
\newcommand{\mfrac}[2]{\mbox{$\frac{#1}{#2}$}}
\newcommand{\pad}[2]{\frac{\partial #1}{\partial #2}}
\newcommand{\padd}[2]{\frac{\partial^2 #1}{\partial #2^2}}
\newcommand{\half}{\mbox{$\frac{1}{2}$}}

\newcommand{\nn}{\nonumber}
\newcommand{\ds}{\displaystyle}
\newcommand{\ep}{\epsilon}
\newcommand{\ol}[1]{\overline{#1}}
\newcommand{\fo}{f_1}
\newcommand{\er}{\langle r \rangle}
\newcommand{\err}{\langle r^2 \rangle}
\let\tilde=\widetilde
\let\hat=\widehat

\newcommand{\BD}{Becker-D\"{o}ring}

\newcommand{\etal}{{\em et al.}}
\newcommand{\ro}{\varrho}
\newcommand{\msum}{\sum_{m=1}^N}
\newcommand{\nsum}{\sum_{n=1}^N}
\newcommand{\psum}{\sum_{p=1}^N}
\newcommand{\qsum}{\sum_{q=1}^N}
\newcommand{\pqsum}{\sum_{p,q=1}^{N,N}}
\newcommand{\mpsum}{\sum_{m,p=1}^{N,N}}
\newcommand{\mpqsum}{\sum_{m,p,q=1}^{N,N,N}}
\newcommand{\rsum}{\sum_{r=1}^{\infty}}
\newcommand{\ssum}{\sum_{s=1}^{\infty}}
\newcommand{\ksum}{\sum_{k=1}^{\infty}}

\newcommand{\chiS}{\chi_{_{\rm\bf S}}}
\newcommand{\chiL}{\chi_{_{\rm\bf L}}}
\newcommand{\chiX}{\chi_{_{\rm\bf X}}}
\newcommand{\epL}{\varepsilon_{_{\rm\bf L}}}
\newcommand{\epS}{\varepsilon_{_{\rm\bf S}}}
\newcommand{\alphaS}{\alpha_{_{\rm\bf S}}}
\newcommand{\alphaL}{\alpha_{_{\rm\bf L}}}
\newcommand{\zetaSS}{\zeta_{_{\rm\bf SS}}}
\newcommand{\zetaSL}{\zeta_{_{\rm\bf SL}}}
\newcommand{\zetaLL}{\zeta_{_{\rm\bf LL}}}

\newcommand{\rominus}[1]{\left(\ro\!\!-\!\!\Lambda NY\!\!-\!\!2\Lambda N^2Z\right)^{#1}}
\newcommand{\epSplus}[1]{\left(\!\epS \!\!+\!\! \alphaS Y\!\!+\!\!\chiS NY\!\!+\!\!\chiX N^2Z\!\!+\!\!\zetaSS N^2YZ\!\!+\!\!2\zetaSL N^3Z^2\!\right)^{#1}}
\newcommand{\epLplus}[1]{\left(\!\epL\!\!+\!\!\alphaL Z\!\!+\!\!\chiL N^2Z\!\!+\!\!\zetaLL N^2Z ^2\!\right)^{#1}}

\newcommand{\lbl}[1]{\label{#1}}
\renewcommand{\baselinestretch}{1.0}

\begin{document}
\renewcommand{\theequation}{\arabic{section}.\arabic{equation}}
\maketitle


\vspace*{-15mm}

\renewcommand{\baselinestretch}{1.3}
\normalsize

\begin{abstract}

\noindent The aims of this paper are to propose, 
construct and analyse microscopic kinetic models 
for the emergence of long chains of RNA from monomeric
$\beta$-D-ribonucleotide precursors in prebiotic 
circumstances.  Our theory starts out from similar 
but more general chemical assumptions to those of 
Eigen \cite{eigen}, namely that catalytic replication 
can lead to a large population of long chains. In 
particular, our models incorporate the possibility 
of (i) direct chain growth, (ii) template-assisted 
synthesis and (iii) catalysis by RNA replicase 
ribozymes, all with varying degrees of efficiency. 
However, in our models the reaction mechanisms are 
kept `open'; we do not assume the existence of closed 
hypercycles which sustain a population of long chains. 
Rather it is the feasibility of the initial emergence 
of a self-sustaining set of RNA chains from monomeric 
nucleotides which is our prime concern.  Moreover, we 
confront directly the central nonlinear features of 
the problem, which have often been overlooked in 
previous studies. Our detailed microscopic kinetic models 
lead to kinetic equations which are generalisations of the 
Becker-D\"{o}ring system for the step-wise growth of 
clusters or polymer chains; they lie within a general
theoretical framework which has recently been successfully 
applied to a wide range of complex chemical problems.  
In fact, the most accurate model we consider has \BD\ 
aggregation terms, together with a general Smoluchowski 
fragmentation term to model the competing hydrolysis of 
RNA polymer chains. We conclude that, starting from 
reasonable initial conditions of monomeric nucleotide 
concentrations within a prebiotic soup and in an 
acceptable timescale, it is possible for a self-replicating 
subset of polyribonucleotide chains to be selected, 
while less efficient replicators become extinct.  

\end{abstract}

\vspace*{-3mm}

\noindent\hrulefill\\
$^*$ corresponding authors \hfill 
\newpage
\scriptsize
\renewcommand{\baselinestretch}{1.5}
\normalsize
\newpage

\section{Introduction}  \setcounter{equation}{0}

Since the early 1970s, an impressive body of 
theoretical and experimental work has been 
accumulating which supports the so-called ``RNA 
world'' view \cite{Gilbert,orgel}.  According to this 
picture,  the central dogma of molecular biology, 
that ``DNA makes RNA makes protein'', is replaced 
with the view that ``in the beginning'' both the genetic 
material and the (enzymatic) catalysts were comprised of
RNA. Altman and Cech won the 1989 Nobel prize in chemistry for 
demonstrating that RNA does indeed have catalytic powers, and that there
are naturally occurring enzymes made of RNA, now known as ribozymes 
\cite{ch95}. According to this standpoint, the problem 
of the evolution of life thus divides 
into two parts: (i) where did the RNA world come from, and (ii) how
did that develop into the world we know today 
involving DNA, RNA and proteins?
Both of these questions are formidable, but whereas the second is widely
appreciated, the first is more easily overlooked \cite{mss}.

This paper aims to address the first of these two questions. When the
issue is considered carefully, it is apparent that it is by no means
straight forward to provide a convincing answer. On the one hand, there is
a lack of evidence for the spontaneous formation 
of the $\beta$-D-ribonucleotide monomers that
comprise RNA under plausible prebiotic conditions (the pyrimidine-containing
compounds in particular have failed to appear in 
all experiments to date); it remains unclear 
how nucleotides of the right chirality
appeared and what led to stereospecific $3^{\prime}-5^{\prime}$
polymerization; and finally no
feasible prebiotic RNA replicator has yet been found. However, 
progress is being made in this direction; for example, it is
known that chemically-activated nucleotides can form oligomeric chains of
20-40 nucleotides in length in the absence of templates \cite{joyce89}. It
seems at least reasonable to be optimistic on this score and to assume that 
in the not too distant future truly autocatalytic RNA molecules will
be made. 

On the other hand, due to the distinct
information content carried by different nucleotide 
base sequences within RNA polymers, the 
problem has a combinatorial complexity which puts it into the class of
NP hard problems in the sense of algorithmic complexity theory
\cite{ch95}. In
particular, because there are four bases within RNA (A, G, C and U), 
a linear RNA polymer of length $N$ monomers can exist as $4^N$ 
chains with different base sequences.
Not only does this preclude numerical analysis of the problem for anything
other than very small values of $N$, it has been used {\it prima facie} as
an implausibility argument against the very possibility of producing
self-replicating RNA molecules from a prebiotic soup comprised
initially of nucleotide monomers. According to this argument, to
produce a significant concentration of any one self-replicating RNA 
molecule--assuming this to be chemically feasible--would 
have required a greater mass of molecules in the prebiotic soup 
than is available from the mass of the Earth~\cite{joy-org}. Such an 
argument may be rendered invalid if the kinetics of the system are
highly nonlinear, as indeed they are in the models we present here.  

Putting aside the several outstanding chemical questions concerning
the realization of actual examples of self-replicating RNAs on
the assumption that these can and will eventually be achieved, the
aim of the present paper is to present and analyse 
kinetic models of 
RNA chain formation and self-replication in prebiotic 
conditions. The central purpose of this work is
to assess the feasibility of achieving viable concentrations of
self-replicating RNA polymers from plausible estimates of 
the physicochemical parameters and conditions which 
most likely pertained under prebiotic conditions.
(Although for the purposes of the study presented 
in this article, only the RNA worldview is considered, 
we wish to point out that there is by no means a 
consensus on this question; for an overview of different
standpoints see Fleischaker \etal\ \cite{fleisch}.)

One part of our paper is concerned with
proposing detailed microscopic kinetic 
models of RNA formation and self-replication; these models 
are loosely based on what we refer to as the so-called 
``\BD" assumptions of classical nucleation theory.  
The major extensions and generalisations of this theory which are needed to 
faithfully model the complexity 
of the current situation lead to equations 
which, as they stand, are far too complex to analyse 
theoretically or numerically. Moreover, the chemistry
which these rate processes describe is far too detailed to be of direct
use to experimentalists. Hence we formulate a ``coarse-graining'' 
reduction aimed at overcoming the combinatorial 
complexity inherent in these systems, and which simplifies the 
kinetic equations so that useful analysis
and comparisons with experiment are possible.  
This reduction procedure is a major theme 
of the present paper.  

In some theoretical respects our work is close to that of Eigen 
\cite{eigen,eigenpop,ems88,eigmccsch,eigenetal} 
and Nu\~{n}o \etal\ \cite{cn,cnp,ne2,ne1,nmr,nt} in that we are 
modelling the formation of RNA as a complex sequence of chemical 
reactions which proceed via catalytic and non-catalytic pathways.  
The main differences between our approach and that of both Nu\~{n}o and Eigen 
are that our reaction networks are all {\em open-ended} rather than
closed cycles (``hypercycles''), and that we address the full
nonlinearity of the kinetic equations involved.   

Moreover, from a chemical perspective 
our work is mainly aimed at answering a significantly different 
question to that posed by Eigen and Nu\~{n}o \etal\ \  Their 
work explains how large concentrations 
of long chains cooperate in order to maintain unexpectedly high 
concentrations when one might expect only very few, if any at all, to be 
present. By contrast, our work deals with an earlier stage of the process, 
namely how long chains are formed from a system which initially has 
only the basic building blocks (monomers) present.  The temporal 
behaviour of our models is, in fact, very similar to that 
shown in figure 16 of Eigen \cite{eigen}.  

Because the full problem we are addressing is mathematically very complex,
we approach it in a sequence of stages within 
the different sections of the paper as now outlined. 
In Section \ref{heuristic}, a highly simplified but instructive
heuristic model is described 
and analysed.  This idealised ``toy'' model shows 
how the various essential rate processes involved, when combined 
together, produce qualitatively the type of behaviour we would hope 
to see in the more realistic and much more complex models of
RNA formation and self-replication we later construct. 

The remainder of the paper presents and analyses various 
more detailed models,  based on the \BD\ equations for 
step-wise growth. The main point to make here is that 
Sections \ref{detail-model-sec} and \ref{cg-sec}
are concerned respectively with detailed and coarse-grained 
models of RNA chain formation and growth in the absence of 
hydrolysis (which breaks up growing chains), whereas 
Sections \ref{det-hydrolysis} and \ref{red-hydrolysis} include 
this important process and demonstrate the dramatic effect it 
has on maximal chain length expected. We then investigate 
under what circumstances some species become extinct and 
others dominate.  

The models derived in Sections \ref{detail-model-sec} 
and \ref{det-hydrolysis} are vastly more complex than 
the original \BD\ equations \cite{bd35,pl76}--and indeed 
even than the generalizations we have previously 
developed to describe micelle and vesicle 
self-reproduction \cite{cw96,cw98}, as well as 
generalised nucleation processes \cite{wc97}--since 
in the present work we attempt to keep track of the 
information contained in the RNA sequences by noting 
the order of addition of nucleotides. This then 
enables us to incorporate the effects of 
template-assisted and ribozymic catalysis of chain 
formation and replication.  The paper closes with a 
discussion in Section \ref{discuss} while an Appendix 
describes a model similar to the one developed in 
Section \ref{detail-model-sec} but incorporating  
more complex kinetic processes.


\section{A simple heuristic model} \label{heuristic}  

In this section we present a very simple heuristic 
model of RNA formation and self-replication which 
contains the basic and essential properties of slow 
growth by an uncatalysed reaction mechanism, fast 
growth due to catalysis and error-prone replication, 
and we show that it produces results consistent with 
our intuition\ --\ namely a long induction time, after 
which large concentrations of products are formed. 

We assume there is a steady source of nucleotides from which to 
build chains; then, with no catalysis and no errors during chain
polymerization (growth), we 
expect an exponential growth in concentration ($u$) as described by 
\beq  \pad{u}{t} = u .   \eeq
Now we shall allow $u$ to depend also on the composition of the 
growing RNA chains (that is, on the sequence of nucleotide bases of
which they are comprised), which 
we denote ${\bf x}$.  Errors in replication of ${\bf x}$ lead to 
the creation of chains with similar composition, and are thus close 
to ${\bf x}$ in this ``sequence space''; this can be thought of as 
a diffusive process in sequence space.  Thus we propose the
following partial differential equation 
\beq
\pad{u}{t} = D\nabla_{({\bf x})}^2 u + u + \alpha u^2 . \lbl{upde}
\eeq
On the right hand side, the $D\nabla_{({\bf x})}^2 u$ 
term describes the diffusive spread of material from 
one type of composition to another via imperfect replication, while the 
term $\alpha u^2$ models catalytic replication, whereby if some 
species $({\bf x})$ exists, its rate of production will depend on 
its own concentration $u({\bf x},t)$. Equation~(\ref{upde}) has the form
of a non-linear reaction-diffusion equation; indeed, equations of
this general structure are well known and their properties have been
much studied in the context of a mathematical theory of explosive
chemical reactions (thermochemical runaway); see, for example, work 
by Dold \cite{dold}. 

Our first step is to show that there is a growing solution 
which is independent of the information in the chains, $u = f(t)$, 
independent of ${\bf x}$.  From equation~(\ref{upde}), 
this amounts to solving the ordinary
differential equation (ODE) 
\beq \dot{f}=f+\alpha f^2 , \lbl{fode} \eeq 
which has a solution $f(t)=f_0 e^t/(1+\alpha f_0(1-e^t))$.   
This expression for $f(t)$ blows up in a finite time, 
at $t_c=\log(1+1/\alpha f_0)$. 

Now we show that the behaviour predicted by this solution is unstable,
in the sense that 
a perturbation of the solution which depends on the information 
variable ${\bf x}$ (i.e., on the nucleotide sequence of the chain)
has an even faster growth rate. 
We put $u({\bf x},t) = f(t) + h({\bf x},t)$, with $h\ll f$. 
We linearise about $f(t)$ and find that $h({\bf x},t)$ satisfies 
\beq
\pad{h}{t} = D\nabla_{({\bf x})}^2 h + h + 
\frac{2 \alpha f_0 e^t h}{1+\alpha f_0 (1-e^t)} , 
\eeq
which can be solved by separation of variables, 
$h({\bf x},t) = X({\bf x}) T(t)$: 
\beq
\frac{T'(t)}{T(t)} - 1 - \frac{2\alpha f_0 e^t}{1+\alpha f_0 (1-e^t)} = 
\frac{D\nabla_{({\bf x})}^2 X({\bf x})}{X({\bf x})} = -\mu . 
\eeq
In the simplest case, where we take a one-dimensional information space 
${\bf x}\equiv x\in{\bf R}^1$, the shape in $x$-space of the variation 
$h({\bf x},t)$ is $\sin(x\sqrt{\mu/D})$, $\cos(x\sqrt{\mu/D})$, indicating 
an increase in the concentration for some $x$ values and a decrease 
in others (relative to the average concentration $f(t)$). 
In fact, this formulation is valid whatever dimension the space of all 
${\bf x}$'s is taken to have.  The temporal evolution of the perturbation 
is unaffected by $X({\bf x})$, and is given by 
\beq
T(t) = \frac{B (1+\alpha f_0)^2 e^{(1-\mu)t}}{(1+\alpha f_0 (1-e^t))^2} . 
\eeq
This function increases more rapidly than $f$ for $t$ sufficiently 
close to $t_c$ (if $\mu>1+2\alpha f_0$ then $h$ at first decreases 
and then increases, blowing up at $t=t_c$). 

In summary, the solution to the heuristic 
equation~(\ref{upde}) we have derived is 
\beq
u({\bf x},t) \sim \frac{f_0 e^t}{1+\alpha f_0(1-e^t)} + 
\frac{B (1+\alpha f_0)^2 X({\bf x}) e^{(1-\mu)t}}{(1+\alpha f_0 (1-e^t))^2}, 
\eeq
which, for times near blow-up ($t\stackrel{<}{\approx} t_c =\log(1+1/\alpha
f_0)$), shows that 
a system with modulations in composition (${\bf x}$) has faster 
growth rates than a system without such dependency, 
indicating that some species will be preferred over others on the
basis of their specific chemical composition (information content).

Although the very simple model presented and analysed in this section
displays some rather unphysical behaviour--such as the presence of
a singularity corresponding to the development of infinite
concentrations of RNA chains in finite time--the very fact that it
is possible to produce very high concentrations of products starting from
arbitrarily low concentrations of reactants is encouraging for our
quest to account for the origin of the RNA world. Our next step is to
construct and analyse the dynamical behaviour of more realistic, and
of course much more complicated, models of RNA formation and
self-replication. It is possible to tame the
singularities which develop in these more realistic models; 
as we shall demonstrate, one means of achieving this is by 
incorporating hydrolysis into such schemes,
which causes the breakdown of longer chains (Sections 
\ref{det-hydrolysis} and \ref{red-hy-sec}). 

\section{Models of RNA polymerization in the absence of hydrolysis}  
\lbl{detail-model-sec}
\setcounter{equation}{0} \label{modelsec}

Our aim now is to begin to model the chemical processes involved in 
the formation and replication of RNA.   We will denote the four  
nucleotides bases $A$, $C$, $G$, $U$ by $N_i$ with $i=1,2,3,4$; 
and oligomeric ribonucleotide 
sequences by $C^\gamma_r$, where $r$ signifies the 
number of bases in the sequence and $\gamma$ denotes the particular
order in which they occur.

The main reactions that such chains can undergo are:
\begin{description}
\item[({\bf i})] the basic \BD\ rate processes controlling chain growth
\beq
C_r^\gamma+N_i \stackrel{{\rm slow}}{\rightleftharpoons}
C_{r+1}^{\gamma+N_i};
\lbl{greac} \eeq
\item[({\bf ii})] template-based chain synthesis (a form of catalysis mediated by
Watson-Crick base-pairing of ribonucleotides on complementary chains) 
\beq
C_r^\gamma + N_i + C_s^\theta \stackrel{{\rm fast}}{\rightleftharpoons} 
C_{r+1}^{\gamma+N_i} + C_s^\theta .  \lbl{creac}
\eeq
Some nucleotide sequences $C_r^\gamma$ are better `replicators' (templates) 
than others, by virtue of base-pairing effects. 
In particular we expect that the complementary chain 
to $\gamma$ (in the sense of Watson-Crick hydrogen-bonded base pairs), 
which we shall denote $\gamma*$, will be an extremely 
effective template for the production of $\gamma$.  
\item[({\bf iii})] `poisoning' or inhibition, for example by tightly-bound duplex
formation
\beq
C_r^\gamma + C_s^\theta \rightleftharpoons P_{r,s}^{\gamma,\theta}.
\lbl{poireac}
\eeq
In the present paper, we have not made any serious attempts 
to incorporate inhibition into our detailed models.
\item[({\bf iv})] hydrolysis -- whereby a long chain is split into two shorter 
chains. Chemically this corresponds to the reaction 
$C_{r+s}^{\gamma+\theta}\rightarrow C_r^\gamma + C_s^\theta$. 
This has the form of a general fragmentation process as modelled 
by the Smoluchowski equations~\cite{Smol}, a mechanism which increases 
the number of chains but reduces the average chain length.  
\item[({\bf v})] enzymatic replication (replicase ribozymal activity) -- 
where a third chain aids the growth 
of a chain which is already in close contact with another chain acting
as a template 
\beq
C_r^\gamma + N_i + C_{r+k}^{\gamma*+\theta*} + C_s^\xi 
\rightleftharpoons 
C_{r+1}^{\gamma+N_i} + C_{r+k}^{\gamma*+\theta*} + C_s^\xi ; 
\eeq 
here the combination of $\gamma$ with $N_i$ is a subsequence of 
the chain $\gamma+\theta$ (that is $\gamma+N_i \subset 
\gamma+\theta$) while $C_s^\xi$ plays the part of a replicase
ribozyme. Needless to say, some of these replicases will have much
higher efficiency than (most) of the others.
\end{description}

We note in passing that, according to Eigen \etal~\cite{eigenetal}, 
the only efficient 
self-replicators have $r \approx 70$ (longer chains have copying 
errors too high to guarantee accurate replication). Our model 
ultimately confirms that there is a critical length, beyond which 
chains cannot self-replicate with sufficient accuracy to remain 
viable.  The manner in which this critical length depends on 
the self-replication error rate, the uncatalysed growth rates
and autocatalytic rates is given in Section \ref{sec62}.

We use $C_r^\gamma$ to denote 
the chain $\gamma$ which has length $r$, and other chains which it may 
interact with will be denoted $C_s^\theta$, whose sequence of bases is 
encoded in $\theta$ and has length $s$. 
We define a flux corresponding to the addition of each type of monomer 
($\{N_i\}_{i=1}^4=\{A,C,G,U\}$) to each type of chain $\gamma$. Since 
there are two ends to each chain there are two growth points. 
The reactions are given by 
\beq
C_r^\gamma + N_i \rightleftharpoons C_{r+1}^{\gamma+N_i} , 
\;\;\;\;{\rm and}\;\;\;\;
N_i + C_r^\gamma \rightleftharpoons C_{r+1}^{N_i+\gamma} .
\eeq
The rates of these reactions will include effects arising from replicase
ribozymal activity, as shown explicitly in the
flux functions $J_r^{\gamma,N}$ below.   The superscripts carry information 
on the exact sequences of bases, and of course the order of the bases is 
important--$N_1$ followed by $\gamma$ is a different sequence 
to $\gamma$ followed by $N_1$.  

We start the mathematical modelling of these processes by considering 
only the reversible aggregation mechanisms.  These all fall into 
the class of \BD\ processes of stepwise cluster growth.  In the first part
of this paper, we shall neglect the hydrolysis and the poisoning 
mechanisms in order to investigate in detail the primary growth 
mechanisms that such an already formidably complex system allows.  
Later on, we shall return to study the effect of hydrolytic fragmentation
mechanisms (Sections~\ref{det-hydrolysis} and \ref{red-hydrolysis}), 
postponing further consideration of inhibition for a subsequent
publication. 

The flux $J_r^{[\gamma,N_i]}$ denotes the rate of attachment 
of nucleotide monomer (base) $N_i$ to the right hand end of $\gamma$.  
Similarly $J_r^{[N_i,\gamma]}$ denotes the rate of attachment 
of monomer $N_i$ to the left hand end of $\gamma$.  
The processes described above mention three mechanisms for 
growth: an uncatalysed chain growth step, two template-based 
growth steps which lead to quadratic catalysis and a ribozymal replicase 
process which is cubic in the reactant concentrations.   We take 
template-based chain growth as being error-prone in that it is most effective 
when $\theta=\gamma*$, but it will also affect the overall 
rate of aggregation when $\theta\neq\gamma*$: 

\beqa \hspace*{-5mm}
J_r^{[\gamma,N_i]} & = & \left( c_r^{[\gamma]} c_1^{[N_i]} - 
\ds \frac{ \ol{c}_r^{[\gamma]} \,
\ol{c}_1^{[N_i]}}{\ol{c}_{r+1}^{[\gamma+N_i]}} 
c_{r+1}^{[\gamma+N_i]} \right) \left( \ep_{r,0}^{[\gamma,N_i]} + 
\alpha_r^{[\gamma,N_i]} c_{r+1}^{[\gamma+N_i]} +
\sum_\theta \chi_{r,s}^{[\gamma,N_i,\theta]} c_s^{[\theta]} + 
\sum_\theta\!\!\sum_{{{\xi:}\atop{\gamma+N_i\subset\xi}}}\!\! 
\zeta_{r,s,k}^{[\gamma,N_i,\theta,\xi]} c_s^{[\theta]} 
c_{r+k}^{[\xi]} \right) , 
\hspace*{-4mm}\nn\\&&\lbl{genfluxa} \\ \hspace*{-5mm}
J_r^{[N_i,\gamma]} & = & \left( c_r^{[\gamma]} c_1^{[N_i]} - 
\ds \frac{ \ol{c}_r^{[\gamma]} \,
\ol{c}_1^{[N_i]}}{\ol{c}_{r+1}^{[N_i+\gamma]}} 
c_{r+1}^{[N_i+\gamma]} \right) \left( \ep_{r,0}^{[N_i,\gamma]} + 
\alpha_r^{[N_i,\gamma]} c_{r+1}^{[N_i+\gamma]} +
\sum_\theta \chi_{r,s}^{[N_i,\gamma,\theta]} c_s^{[\theta]} + 
\sum_\theta \!\!\sum_{{{\xi:}\atop{N_i+\gamma\subset\xi}}}\!\! 
\zeta_{r,s,k}^{[N_i,\gamma,\theta,\xi]} c_s^{[\theta]} 
c_{k+r}^{[\xi]} \right) .   \hspace*{-4mm}\nn\\&&\lbl{genfluxb}
\eeqa
We use $\gamma_l$ to denote the left hand end monomer of 
the chain $\gamma$, and $\gamma_r$ for the right end monomer;  
$\gamma-\gamma_r$ denotes the chain $\gamma$ without the 
final ($\gamma_r$) monomer, and $-\gamma_l+\gamma$ the 
chain without its first element.  When this information is included 
in the notation of a concentration or a rate constant, we shall place 
it as a superscript in square brackets to avoid any possible 
confusion with exponents.  Using this notation, the kinetic 
equation for the concentration of the chain $\gamma$ is then 
\beqa 
\dot{c}_r^{[\gamma]} & = & J_{r-1}^{[\gamma_l,-\gamma_l+\gamma]} + 
J_{r-1}^{[\gamma-\gamma_r,\gamma_r]} - \ds \sum_{i=1}^4 J_r^{[N_i,\gamma]}
- \ds \sum_{i=1}^4 J_r^{[\gamma,N_i]} \lbl{genkin} \\ && \nn
- \ds \sum_\theta k_{r,s}^{[\gamma,\theta]} \, c_r^{[\gamma]}\,c_s^{[\theta]} 
- \ds \sum_\theta k_{s,r}^{[\theta,\gamma]} \, c_r^{[\gamma]}\,c_s^{[\theta]} 
+ \ds \sum_\theta m_{r,s}^{[\gamma,\theta]} \, p_{r,s}^{[\gamma,\theta]} 
+ \ds \sum_\theta m_{s,r}^{[\theta,\gamma]} \, p_{s,r}^{[\theta,\gamma]} 
\eeqa
(the apparent duplication of terms being due to the assumption that 
the model must be able to distinguish which end of the chain a 
monomer is added to).  

These equations introduce the poisoned duplices $P_{r,s}^{\gamma,\theta}$ 
whose concentrations $p_{r,s}^{[\gamma,\theta]}$ satisfy the equation 
\beq
\dot{p}_{r,s}^{[\gamma,\theta]} = k_{r,s}^{[\gamma,\theta]}\,c_r^{[\gamma]} 
\, c_s^{[\theta]} - m_{r,s}^{[\gamma,\theta]} \, p_{r,s}^{[\gamma,\theta]} . 
\lbl{peq}\eeq
In constructing these equations, we have assumed the existence of 
a unique equilibrium solution for the $c_r^{[\gamma]}$'s, written
$\ol{c}_r^{[\gamma]}$.  Provided this exists, an equilibrium solution 
for the poisoned duplices $p_{r,s}^{[\gamma,\theta]}$ can be found 
trivially from (\ref{peq}) 
\beq
\ol{p}_{r,s}^{[\gamma,\theta]} = 
\frac{ k_{r,s}^{[\gamma,\theta]} \: \ol{c}_r^{[\gamma]} \:
\ol{c}_s^{[\theta]} }{ m_{r,s}^{[\gamma,\theta]} } . 
\eeq
The other main assumption we have made here is that all four types of monomer 
are held at a constant concentration, the so-called pool chemical
approximation familiar in chemical kinetics.  

In the remainder of this paper, we shall ignore 
poisoning altogether; this can be thought of as 
equivalent to setting all the coefficients 
$k_{r,s}^{[\gamma,\theta]}$, $m_{r,s}^{[\gamma,\theta]}$ 
equal to zero. As written above, these models are 
extremely complex and indeed represent problems that 
are of NP type in that their algorithmic `size' is an 
exponential function of the chain length; this length 
may itself become unbounded as the reactions proceed.  
Such intractability means that any analysis must be 
performed on drastic simplifications of these kinetic 
equations. 

Even these kinetic equations are of course idealised 
and no experimental systems have yet been identified 
which exactly realise the steps we are proposing. 
For example, such polymerisations normally need activating 
ester linkages; and the polycondensation of mononucleotides 
diluted in water yields water as a byproduct--thus reducing 
the concentration of all species.  However it is quite 
feasible that other, more appropriate chemistries will 
be discovered in the not too distant future. 

We shall now analyse two simplified models, one of which 
exhibits only autocatalysis on the microscopic scale, 
while the other is fully crosscatalytic.  

To simplify the models we shall assume that there is a significant 
autocatalytic mechanism (specifically we assume that the concentration 
of a chain $c_r^{[\gamma]}$ is approximately equal to that of its 
complement $c_r^{[\gamma*]})$, that all chains have the {\em same} 
small crosscatalytic effect, and that the role played by the
replicase ribozymes is both 
crosscatalytic and autocatalytic.   Thus we shall replace the vast 
number of possible but generally unknown 
rate coefficients in (\ref{genfluxa})--(\ref{genfluxb}) 
by just four rate constants $\ep,\alpha,\chi,\zeta$, an approximation 
which simplifies 
the flux functions occurring in those equations to 
\beq
J_r^{[\gamma,N_i]} = \left( c_r^{[\gamma]} c_1^{[N_i]} - 
\ds \frac{ \ol{c}_r^{[\gamma]} \,
\ol{c}_1^{[N_i]}}{\ol{c}_{r+1}^{[N_i+\gamma]}} 
c_{r+1}^{[N_i+\gamma]} \right) \left( \ep + \alpha c_{r+1}^{[\gamma+N_i]} + 
\chi \sum_\theta c_s^{[\theta]} + \zeta c_{r+1}^{[\gamma+N_i]} 
\sum_\theta c_s^{[\theta]} \right) .  \lbl{simplerJ}
\eeq


\subsection{Analysis of RNA formation for the case of pure autocatalysis}  
\lbl{auto-sec}

Next we neglect the breakup of chains, so that (\ref{greac}) and (\ref{creac}) 
are treated as irreversible; we also ignore the replicase mechanism 
and template crosscatalysis (that is, we neglect error-prone 
template synthesis). 
The resulting equations are made manageable if we 
analyse a case of this model in which chains catalyse only their own 
production--thus a chain $c_{r+1}^{\gamma+N_3}$ can be constructed 
in two ways: 
\beq
C_r^\gamma + C_1^{N_3} \stackrel{\ep}{\longrightarrow} 
C_{r+1}^{\gamma+N_3} , \hspace{7mm} C_r^\gamma + C_1^{N_3} + 
C_{r+1}^{\gamma+N_3} \stackrel{\alpha}{\longrightarrow} 2 
C_{r+1}^{\gamma+N_3} ,
\eeq
where $\alpha$ and $\ep$ denote rate coefficients.
By generalising $\alpha$ to $\alpha+\beta r$ ($r$ being the length of
the chain),  we can consider a system 
where longer chains can have greater catalytic effect than short chains.
The fluxes are then defined by 
\beq 
J_r^{\gamma,N_i} = c_r^\gamma c_1^{N_i} \left( \ep
+ (\alpha+\beta r) c_{r+1}^{\gamma+N_i} \right) , \hspace{5mm} 
J_r^{N_i,\gamma} = c_r^\gamma c_1^{N_i} \left( \ep 
+ (\alpha+\beta r) c_{r+1}^{N_i+\gamma} \right) , \lbl{acJ}
\eeq
with kinetics determined by 
\beq
\dot c_r^\gamma = J_{r-1}^{\gamma_1,-\gamma_1+\gamma} + 
J_{r-1}^{\gamma-\gamma_r,\gamma_r} - \sum_{i=1}^4 J_r^{\gamma,N_i} - 
\sum_{i=1}^4 J_r^{N_i,\gamma} .  \lbl{ackin1}
\eeq
We study the system of equations in the pool chemical approximation by
assuming a constant concentration of 
monomers, so that the total matter density will {\em not} be preserved.   
We now proceed to solve the \BD\ equations and show that 
growth of long chains can indeed occur. 

We seek a solution assuming that the rate coefficients are of
magnitudes $\ep\ll1$, $\alpha,\beta = {\cal O}(1)$ (that is, 
the uncatalysed rate is much smaller than the template 
catalysed rate) and with concentrations depending only on chain 
length ($r$). From equations (\ref{acJ}) and (\ref{ackin1}), we find that 
\beq
\dot{c}_r=-8 c_r c_1 \left( \ep+(\alpha\!+\!\beta r) c_{r+1} \right) + 
2 c_1 c_{r-1} \left( \ep + (\alpha\!+\!\beta(r-1)) c_r \right) .
\eeq 
Strictly speaking, since the equations model irreversible chain 
growth, the system cannot be described as having an equilibrium 
solution in the thermodynamic sense; instead,  it approaches a 
steady state solution.  In the case $\ep\rightarrow0$, this is 
\beq
c_r = \frac{ 2^{1-r}\,\ol{c}_1\,
\Gamma(\half (r\!+\!\alpha/\beta))\;\Gamma(\half(2\!+\!\alpha/\beta))} 
{\Gamma(\half(r\!+\!1\!+\!\alpha/\beta))\;\Gamma(\half(1\!+\!\alpha/\beta))}.
\eeq

Increasing the catalytic effect of longer chains produces a more rapidly 
decaying distribution of long chains.  This is perhaps slightly 
counter-intuitive: we have made it easier to make long chains and yet 
find fewer of them; the apparent paradox is resolved by 
noting that long chains are less 
likely to `hang around' as they are busy forming even longer chains. 
In more realistic models there would be fewer long chains due 
to the larger errors encountered in replicating long chains, not to
mention the effect of hydrolysis. 


\subsection{Analysis of error-prone RNA formation} 
\lbl{cross-sec}

The above analysis showed, using a very simple model, 
that autocatalysis alone could produce a population of long chains.
We now turn to crosscatalysis to see if this also aids the 
production of longer chains.  Here we are obliged to analyse a more complex
model, where all chains act as catalysts for growth.
In what follows, we shall analyse the simplest case, for which the 
rate constants are 
independent of chain length; in Appendix~\ref{appa} we analyse a  
more complex case where the rate constants are proportional to chain
length (and for which more singular behaviour is found).  

In this case the rate coefficients are given by 
$\alpha_{r,s}^{\gamma,\theta}=1$, 
$\alpha_{r,0}^{\gamma,\theta}=\ep$, so that the fluxes can be written as 
\beq
J_r^{N_i,\gamma} = J_r^{\gamma,N_i} = 
c_r^\gamma c_1^{N_i}\left(\ep+\sum_\theta c_s^\theta\right),
\lbl{ccJind} \eeq
in place of (\ref{acJ}).  We continue to use equation (\ref{ackin1}) 
to relate these fluxes to the rate of change of concentration. 
We shall use $f_r = \sum_{\gamma:|\gamma|=r} c_r^\gamma$, where the 
notation $|\gamma|$ means the length of the chain $\gamma$.
The total monomer concentration of the monomer species is denoted 
$\fo = \sum_{i=1}^4 c_1^{N_i}$ and is held fixed to the same constant 
for each of the four types of base ($c_1^{N_i}=\rec{4} \fo \;\; \forall i$). 

The form of fluxes (\ref{ccJind}) implicitly assumes 
that all the rate coefficients are independent of 
$\gamma,N_i,\theta$; thus we look for solutions which 
only depend on the length of the chain and are independent 
of its exact composition. We then obtain the simpler equation
\beq
\dot f_r = 2 \fo f_{r-1} \left(\ep + \sum_{s=1}^\infty f_s \right) - 
2 \fo f_r \left( \ep + \sum_{s=0}^\infty f_s \right) . \lbl{simplereqn}
\eeq
To solve this type of equation, we define the {\em generating function}
\beq
F(z,t) = \rsum f_r(t) e^{-r z} ,  \lbl{Fdef}
\eeq
which, from eqn(~\ref{simplereqn}), satisfies the partial differential equation
\beq
\pad{F(z,t)}{t} = 2 \fo \left[ \ep + F(0,t) \right] 
\left[ \fo e^{-z} - (1-e^{-z}) F(z,t) \right] . 
\lbl{easyPDE}
\eeq

To solve eqn~(\ref{easyPDE}) we first find $F(0,t)$ by letting $z\rightarrow0$ 
in equation (\ref{easyPDE}) to give 
$\mfrac{d}{dt}F(0,t) = 2 \fo^2 [ \ep + F(0,t) ]$ whose solution is 
\beq
F(0,t) = (\fo+\ep) e^{2 \fo^2 t} - \ep . \lbl{crossF0}
\eeq
Now that $F(0,t)$ is known, the differential equation 
(\ref{easyPDE}) can be solved to yield
\beq
F(z,t) = \left( \frac{\fo e^{-z}}{1-e^{-z}} \right) \left[ 1 - \exp \left\{ 
-z - (1-e^{-z}) (1+\ep/\fo) \left( e^{2\fo^2 t} - 1 \right) \right\} \right]. 
\eeq

Ideally we would proceed to find $f_r(t)$ from this 
by forming the Taylor series in $z$, but it turns out that 
there is no simple formula for the $r^{{\rm th}}$ derivative. 
Instead we interpret the chain length distribution function $f_r(t)/F(0,t)$ as 
a time-dependent probability density function; it is then straightforward to 
find how the expected chain length and standard deviation vary with time. 
Firstly, the total number of chains ${\bf N}(t) = F(0,t)$ and so is given
by (\ref{crossF0}).  The total number of chains increases exponentially
starting from $\fo$ at $t=0$. The expected length of chains grows with 
the same exponent:
\beq
{\bf E}(t) \;=\; \ds\frac{\er}{{\bf N}(t)} \; = \; 
\frac{-1}{F(0,t)} \left. \ds\pad{F}{z} \right|_{z=0} \; = \; 
\frac{(\fo+\ep)e^{4t\fo^2}+2(\fo^2-\ep^2)e^{2t\fo^2}-\fo^2-2\fo\ep+\ep^2}
     { 2 \fo ( (\fo+\ep) e^{2t\fo^2} - \ep ) } . 
\eeq
Finally the standard deviation ($\sigma=\sqrt{{\bf V}}$) starts from zero
and also grows with the same exponent, so that in the large-time limit 
the standard deviation has the same order of magnitude as the mean 
chain length:  
\beqa
{\bf V}(t) & = & \ds\frac{\err}{{\bf N}} - \ds\frac{\er^2}{{\bf N}^2} \; =
\;
\ds\frac{1}{F(0,t)} \left. \ds\padd{F}{z} \right|_{z=0} - \left( 
\ds\frac{-1}{F(0,t)} \left. \ds\pad{F}{z} \right|_{z=0} \right)^2 \nn \\ 
& \sim & \rec{12} \left( e^{2t\fo^2} - 1 \right) \left( 
e^{2t\fo^2} + 7 + e^{-2t\fo^2} +3 e^{-4t\fo^2} \right) + {\cal O}(\ep) . 
\eeqa
None of these quantities blows up in finite time, but they all tend to infinity
as $t\rightarrow\infty$. The standard deviation ($\sqrt{{\bf V}(t)}$) and the 
mean (${\bf E}(t)$) both tend to infinity exponentially with the same
exponent. Thus after a large time there will be a wide variety of chain
lengths present in the system. 
 
\subsection{Discussion}

The model analysed in this section allows for the growth of long chains. 
We have shown that these kinds of \BD\ based models 
can exhibit a wide variety of 
behaviours, from the purely autocatalytic model of Section~\ref{auto-sec}
which approaches 
a steady-state solution, through the steady slow growth in length 
and number of chains which occurs in the purely crosscatalytic 
template-based growth model of Section~\ref{cross-sec}, 
to the singular behaviour described in 
Appendix~\ref{appa} for an alternative 
crosscatalytic template-based growth model where rate constants are 
proportional to chain length.  In this last case, the number of chains 
and expected chain length both blow up in finite time.  This behaviour 
would be moderated in the presence of hydrolysis (see Sections 
\ref{det-hydrolysis} and \ref{red-hy-sec}). 

In the next section, we describe a way of analysing 
the behaviour of replicating RNA systems which allows 
different chain compositions of equal length to have 
varying concentrations. This will enable us to 
investigate whether some species (that is, RNA 
sequences) dominate over others and the possibility 
that certain species may become extinct. 

\section{Coarse-grained reductions of microscopic models}  \lbl{cg-sec}
\setcounter{equation}{0}

In the previous section we have considered two massive simplifications to 
the model proposed in Section \ref{modelsec} neither of which was 
capable of describing the effects of the enzymatic reaction mechanism.  
So far, all the solutions we have managed to find have been for situations 
where all chains of the same length have the same concentration 
regardless of their composition.  These solutions have shown the 
dramatic effects, both qualitative and quantitative, which
template-based and ribozymic catalysis can have on the kinetics of growth. 
 
In this section we will perform an alternative 
simplification which enables us to construct 
solutions which allow different chain compositions 
to have different concentrations. The main 
difference between the analysis of the previous two 
sections and this one is that here we will 
initially be returning to reversible reactions 
(although later we shall again impose 
irreversibility). We thus return to equations 
(\ref{genfluxa})--(\ref{genkin}), although as 
noted earlier we shall ignore the inhibition steps.  

\subsection{Contraction} \lbl{con-sec}

The equations we start with are (\ref{genfluxa})--(\ref{genkin}).  
Our aim is to perform a coarse-grained rescaling to reduce the 
number of equations in the system in the same way as has been 
performed in micelle formation \cite{cw96}, vesicle formation 
\cite{cw98}, and generalised nucleation theory \cite{wc97}.   
The same procedure applied here to 
equations~(\ref{genfluxa})--(\ref{genkin}) yields 
\beqa
\dot{x}_r^{[\gamma]} & = & 
L_{r-1}^{[\gamma - \gamma_{r-\lambda}^r, \gamma_{r-\lambda}^r]} + 
L_{r-1}^{[\gamma_1^\lambda, -\gamma_1^\lambda+\gamma]} - 
\sum_{\omega:|\omega|=\lambda}  \left( L_r^{[\gamma,\omega]} + 
L_r^{[\omega,\gamma]} \right) ,  \lbl{cfkin} \\ 
L_r^{[\gamma,\omega]} & = & \left( x_r^{[\gamma]} \ol{x}_1^\lambda - 
\frac{Q_{\Lambda_r}^{[\gamma]}}{Q_{\Lambda_{r+1}}^{[\gamma+\omega]}} 
x_{r+1}^{[\gamma+\omega]} \right) \left( \tilde{\ep}_r^{[\gamma,\omega]} 
+ \tilde{\alpha}_{r+1}^{[\gamma,\omega]} x_{r+1}^{[\gamma+\omega]} + 
\sum_\theta \tilde{\chi}_{r,s}^{[\gamma,\omega,\theta]} x_s^{[\theta]} + 
\sum_\theta \sum_{{{\xi:}\atop{\gamma+\omega\subset\xi}}} 
\tilde{\zeta}_{r,s,k}^{[\gamma,\omega,\theta,\xi]} x_s^{[\theta]} 
x_{k+r}^{[\xi]} \right)^\lambda ,  \nn\\&& \\
L_r^{[\omega,\gamma]} & = & \left( x_r^{[\gamma]} \ol{x}_1^\lambda - 
\frac{Q_{\Lambda_r}^{[\gamma]}}{Q_{\Lambda_{r+1}}^{[\omega+\gamma]}} 
x_{r+1}^{[\omega+\gamma]} \right) \left( \tilde{\ep}_r^{[\omega,\gamma]} 
+ \tilde{\alpha}_{r+1}^{[\omega,\gamma]} x_{r+1}^{[\omega+\gamma]} + 
\sum_\theta \tilde{\chi}_{r,s}^{[\omega,\gamma,\theta]} x_s^{[\theta]} + 
\sum_\theta \sum_{{{\xi:}\atop{\omega+\gamma\subset\xi}}} 
\tilde{\zeta}_{r,s,k}^{[\omega,\gamma,\theta,\xi]} x_s^{[\theta]} 
x_{k+r}^{[\xi]} \right)^\lambda . \nn\\&& \lbl{cfflux} 
\eeqa
Note that the \BD\ structure~\cite{bcp86} 
present in the original system of rate equations is not 
destroyed  by this approximation \cite{cw96}, \cite{cw98},
\cite{wc97}.  Here $\omega$ is an arbitrary 
sequence of length $\lambda$, $\gamma_l^\lambda$ 
represents the sequence of the first $\lambda$ nucleotide monomers of 
$\gamma$, and $\gamma_{r-\lambda}^r$ the last $\lambda$. 

It is inevitable that this coarse-grained 
contraction procedure hides some of the subtleties 
of notation: for example, the term $x_1^\lambda$ 
in the contracted flux does not determine the order 
in which the nucleotides were added--there is no 
longer a unique form for $\gamma$. Thus the 
contraction procedure inherently performs a 
limited averaging in $\gamma$-space (sequence space). 

As an {\em a posteriori} check that these procedures are sensible, 
we average this system over all possible chain compositions (a 
``$\gamma$-average'' of the system (\ref{cfkin})--(\ref{cfflux})).  
This gives exactly the same results as if we had taken a 
$\gamma$-average of the original system 
(\ref{genfluxa})--(\ref{genkin}) 
and then performed the contraction procedure: {\em the processes of 
$\gamma$-averaging and coarse-graining are commutative}.  
Both end up yielding 
\beqa \hspace*{-5mm}
\dot x_r = 2 L_{r-1} - 2 (4^\lambda) L_r,  &\hspace{2mm}& 
L_r  = \left( x_r x_1^\lambda -  \frac{Q_{\Lambda_r}} 
{Q_{\Lambda_{r+1}}} x_{r+1} \right) \left( \tilde{\ep}_r + 
\tilde{\alpha}_{r+1} x_{r+1} + \sum_{s=1}^\infty 
\tilde{\chi}_{r,s} x_s + \sum_{s=1}^\infty \sum_{k=1}^\infty 
\tilde{\zeta}_{r,s,k} x_s x_{r+k} \right)^\lambda .  
\hspace*{-6mm}\nn\\&&\eeqa
Below we shall flesh out in more detail two of 
the simplest approximate schemes which this method 
produces. To aid the clarity of the resulting 
models we shall introduce some new notation 
and drop the tildes from the rate constants, 
on the understanding that they have been rescaled 
in the contraction procedure. 

\subsection{Maximal contraction of the rate processes} \lbl{max-con-sec}

In this section we take the contraction procedure described in the 
previous section as far as possible.  We take a value of $\lambda$ 
large enough so that there are no chains of length $2\lambda$. 
Thus we consider only two types of sequence: the monomer forms 
$x_1$ and long chains with composition $\gamma$, $x_2^\gamma$.  

Consider a combination of crosscatalysis and 
auto-catalysis (that is template-based chain 
synthesis with errors), enzymatic replication,  
as well as the normal reaction mechanism; then 
there are four distinct ways to make a chain, 
described in Table \ref{4tab}. 

\begin{table}[bt]
\[  \begin{array}{rclcl} &&&&\\[1ex]
\Lambda X_1              & \rightleftharpoons &   X_2^\gamma 
&& \mbox{uncatalysed, with forward-rate coeff.} = \ep \\
\Lambda X_1 + X_2^\gamma & \rightleftharpoons & 2 X_2^\gamma
&& \mbox{autocatalysis, with forward-rate coeff.} = \alpha \\ 
\Lambda X_1 + X_2^\theta   & \rightleftharpoons & 
X_2^\gamma + X_2^\theta
&& \mbox{crosscatalysis, with forward-rate coeff.} = \chi  \\
\Lambda X_1 + X_2^\gamma + X_2^\theta   
& \rightleftharpoons & 2 X_2^\gamma + X_2^\theta
&& \mbox{enzymatic-catalysis, with forward-rate coeff.} =\zeta . 
\end{array}  \]
\caption{{\bf The four mechanisms by which long chains 
are formed.}}
\lbl{4tab}
\end{table}

To simplify the notation, since we are only dealing with 
one length of chain ($|\gamma|=\Lambda=\lambda+1$), 
we shall number the types of chain, that is use $Y_n$ in place 
of $X_2^\gamma$ and $y_n$ in place of $x_2^{[\gamma]}$ 
for the corresponding concentrations. Then the equation 
for the rate of change of concentration $y_n$ is 
\beq
\dot y_n = 2 L = 2 \left( x_1^\Lambda - \beta y_n \right) 
\left( \ep + \alpha y_n + \chi \sum_{p=1}^N y_p + 
\zeta y_n \sum_{p=1}^N y_p \right)^\lambda , \lbl{yeq} 
\eeq
where $N=4^\Lambda$ is the number of different chains 
($\beta$-D-ribonucleotide sequences) of length $\Lambda$. 
The parameter $\beta$ determines the ratio of chains 
to monomers at equilibrium. The mechanisms listed 
in Table \ref{4tab} are assumed to leave this ratio  
unchanged; the rates $\ep,\alpha,\chi,\zeta$ only 
alter the timescale over which the system reaches 
equilibrium.  The backward rates for the mechanisms 
listed in Table \ref{4tab} are thus $\beta\ep, 
\beta\alpha, \beta\chi, \beta\zeta$ respectively. 

The advantage of this model equation is that it can be 
solved in a number of different cases, and it is possible 
to find how an initially uneven distribution of ribonucleotide chains 
evolves in time -- for instance whether certain chains die out, or 
if error-prone template synthesis (crosscatalysis) 
will even out the differences.  The 
disadvantage is that the model does not contain any 
information about whether chains will tend to grow 
longer, or how expected chain length varies with time.  

\subsection{Analysis of maximally contracted model}
\lbl{mc-thry-sec}

There are two cases where we shall perform some 
simple analysis to show that the reduced model 
derived above generates meaningful results.  
We shall consider a uniformly growing solution -- 
where all types of chain are present at the same 
concentration levels -- and show that the concentrations 
increase.  We shall then show that this solution is 
unstable to perturbations which favour 
the growth of one species over another.  
The two cases we consider are firstly constant monomer 
concentration, and then constant density when the backwards rates can
be neglected ($\beta = 0$). 

\subsubsection{Prebiotic soup with constant 
concentration of ribonucleotide monomers}

We first consider a prebiotic soup which is provided with a constant 
supply of $\beta$-D-ribonucleotide monomers. 
For simplicity we shall specify the monomer concentration $\ol{x}_1$ 
in such a way that $2\ol{x}_1^\Lambda=1$.  If we then seek a solution 
of the form $y_n(t) =Y(t)$, independent of $n$, we obtain 
\beq
\dot Y = ( \ep + \alpha Y + \chi N Y + \zeta N Y^2 )^\lambda .  
\eeq
In order to solve this with the initial conditions $Y(0)=0$,  we make use 
of asymptotics with $\alpha,\zeta,\chi\sim{\cal O}(1)\gg\ep$.  The reaction 
starts on a timescale of $t_1 = \ep^{\lambda-1}t$, where $Y=\ep Y_1$ 
satisfies $Y'_1=(1\!+\!\alpha Y_1\!+\!N\chi Y_1)^\lambda $, hence 
\beq
Y = \frac{\ep}{(\alpha+N\chi)} \left( 
\frac{1} {(1-t/t_c)^{1/(\lambda-1)}}-1\right) ,
\hspace{9mm} {\rm where} \;\;\;\; 
t_c = \frac{\ep^{1-\lambda}}{(\lambda-1)(\alpha+N\chi)} . 
\lbl{mcx1tc} \eeq
This solution is valid until $Y$ reaches ${\cal O}(1)$, 
when the enzymatic reaction mechanism becomes significant 
and causes a slightly more rapid blow up. 

In practise however, we do not expect to find all species of RNA 
chain present in equal quantities.  Hence in the real world 
this uniform solution must be unstable.  We shall now show 
that it is also unstable in our model, and hence that our 
model does indeed predict the preferential growth of one 
chain type over the others. 

To analyse the stability of the uniform solution, we 
introduce a perturbation around it. Substituting 
$y_n=Y(1+\hat y_n)$ into (\ref{yeq}) and linearising we find
\beq
Y \dot{\hat{y}}_n = \lambda Y 
( \alpha\hat y_n+N\zeta Y\hat y_n+\chi\hat\sigma+\zeta Y \hat\sigma ) 
( \ep + \alpha Y + N \chi Y + \zeta N Y^2 )^{\lambda-1} - 
\hat y_n (  \ep + \alpha Y + N \chi Y + \zeta N Y^2 )^\lambda , 
\eeq
where $\hat\sigma=\nsum \hat y_n$.
Summing the above equation over the index $n$, it is possible to show 
that if $\hat\sigma(t_0)=0$ then $\hat\sigma(t)=0$ 
for all $t$.   This enables the above equation to be simplified to 
\beq
\dot{\hat{y}}_n = \frac{ \hat{y}_n B^{\lambda-1}}{Y} \, \left[ \, 
\lambda Y (\alpha + \zeta N Y) - B \,\right] \, , 
\eeq
where $B = \ep + \alpha Y + \chi N Y + \zeta N Y^2$. 
Thus the uniform solution is unstable whenever the 
difference in the square brackets is positive. 
For small times the uniform solution is stable, 
but at larger times it becomes unstable. 
If $\alpha<N\chi/(\lambda\!-\!1)$ then 
the instability does not set in until $t\sim t_c$, 
but if $\alpha>N\chi/(\lambda\!-\!1)$ then 
the instability occurs for times after 
\beq
t_{{\rm instab}} = t_c \left\{ 1 - \left[ 1 - \left( 
\frac{\alpha+N\chi}{\alpha\lambda} 
\right) \right]^{\lambda-1} \right\} , \lbl{tinstab}
\eeq
that is, once the uniform solution has reached the size 
$Y=Y_{{\rm instab}}=\ep / [\alpha(\lambda\!-\!1)\!-\!N\chi]]$. 
So we see that the instability occurs when the ratio of 
autocatalysis to crosscatalysis (equivalently, the ratio of
error-free to error-prone template synthesis) exceeds a certain 
threshold.  Put another way, {\em for certain species (sequences) 
to grow whilst others 
become extinct, the template-based copying of RNA needs 
to achieve a certain specified level of accuracy.} 

\subsubsection{Prebiotic soup with constant 
total mass of ribonucleotide bases}

In the case of a prebiotic soup with a constant 
total quantity of ribonucleotide bases (another 
way of putting this is to say that the total
nucleotide mass density, which may be present 
in monomers or chains, is constant), we 
eliminate the monomer concentration to leave 
\beq
\dot y_n = 2 \left( \ro - \Lambda \nsum y_n \right)^\Lambda 
\left( \ep + \alpha y_n + \chi \psum y_p + \zeta y_n \psum y_p
\right)^\lambda . 
\eeq
To find the uniform solution and determine its stability, 
we once again use the substitution $y_n = Y ( 1 + \hat y_n) $.  
Inserted into the above, this yields
\beqa
\dot Y & = & 2 ( \ro - \Lambda N Y )^\Lambda 
( \ep + \alpha Y + \chi N Y + \zeta N Y^2 )^\lambda  \lbl{mcroYeq} \\ 
\dot{\hat{y}}_n & = & \frac{ 2\hat{y}_n}{Y} 
(\ro - \Lambda N Y )^\Lambda
( \ep \!+\! \alpha Y \!+\! \chi N Y \!+\! \zeta N Y^2 )^{\lambda-1}
\left[ \lambda (\alpha Y \!+\! \zeta Y^2 N ) - 
( \ep \!+\! \alpha Y \!+\! N \chi Y \!+\! N \zeta Y^2 ) \right] , 
\nn\\&&
\eeqa
where we have already made use of the fact 
that $\hat\sigma=\nsum \hat y_n=0$ for all time. 
The leading order solution to the former (\ref{mcroYeq}) 
is similar to the uniform solution in the constant monomer 
case above: 
\beq
Y = \frac{\ep}{(\alpha+N\chi)} \left( 
\frac{1} {(1-t/t_c)^{1/(\lambda-1)}}-1\right) ,
\hspace{9mm} {\rm where} \;\;\;\; 
t_c = \frac{\ep^{1-\lambda}}{2\ro^\Lambda (\lambda-1)(\alpha+N\chi)} . 
\lbl{mcrotc} \eeq
The uniform solution is then unstable for $t>t_{{\rm instab}}$, 
where $t_{{\rm instab}}$ is given by (\ref{tinstab}) but with 
$t_c$ determined by (\ref{mcrotc}) rather than (\ref{mcx1tc}). 

\subsubsection{Remarks}

It is possible to prove similar results for the 
case in which the backward rates are not neglected ($\beta\neq 0$), 
since all the analysis presented 
in the above section relies on a solution for which $Y$ is small 
($Y\sim{\cal O}(\ep)$). The solution remains valid 
until $Y\sim {\cal O}(1)$.   Since all the instabilities 
found above arise whilst $Y$ is still ${\cal O}(\ep)$, 
they will also arise in systems with $\beta\neq0$, 
because in such systems the uniform solution and its 
stability criteria will not be altered by the $\beta$ 
term until $Y$ reaches ${\cal O}(1)$. 

A more general notion of stability that we have not 
addressed is that for which the rate coefficients differ for 
different ribonucleotide chain sequences.  By this we mean, for 
example, that different chains $y_n$ will have 
different autocatalytic rate constants, $\alpha_n$.
A more detailed analysis could be carried out, and similar 
results to those presented above would be produced. 
The main difference is in the selection of which chains 
proliferate.  In the case studied here, this is 
determined by initial conditions, but for cases in 
which the autocatalytic rate varies with chain composition, 
this provides another selection mechanism and those 
chains with the largest autocatalytic rates 
will grow at the expense of the other species. 
Thus perturbations to the autocatalytic rates 
accentuate the instability analysed above. 

\subsection{Almost maximal contraction}
\lbl{amc-subsec}

In the foregoing work, we have used a maximally-contracted \BD\ model 
to analyse the kinetics of formation of RNA chains.   In order 
to derive a slightly more accurate model we can use the same 
theory as described in Section \ref{con-sec} but instead of 
using a coarse-graining mesh so large that only monomers 
and chains are captured, we use a finer-grained mesh which includes 
two types of chains: short ones (of length $\Lambda$), and long ones 
(of length $\Lambda+\mu$).  Formally, the kinetic equations 
are then 

\beqa
\dot{x}_2^{[\gamma]} & = & L_1^{[\gamma_1^1,\gamma_2^\Lambda]} 
+ L_1^{[\gamma_1^\lambda,\gamma_\Lambda^\Lambda]} - 
\sum_{\omega:|\omega|=\mu} \left( L_2^{[\gamma,\omega]} +
L_2^{[\omega,\gamma]} \right) \nn\\ 
\dot{x}_3^{[\gamma]} & = & 
L_2^{[\gamma_1^\Lambda,\gamma_{\Lambda+1}^{\Lambda+\mu}]} + 
L_2^{[\gamma_1^\mu,\gamma_{\mu+1}^{\Lambda+\mu}]} \\ 
L_1^{[N,\gamma]} & = & \left( x_1^\Lambda - 
\beta_2 x_2^{[N+\gamma]} \right) \left( \ep + \alpha_2 x_2^{[N+\gamma]} + 
\chi \sum_\theta x_2^{[\theta]} + \chi \sum_\theta x_3^{[\theta]}  
+  x_2^{[N+\gamma]} \sum_\theta \zeta x_2^{[\theta]}  
+  x_2^{[N+\gamma]} \sum_\theta \zeta x_3^{[\theta]}  
+ \right. \nn\\&&\nn \left. \hspace*{20mm}
+  \sum_{\theta,\omega} \zeta x_3^{[N+\gamma+\omega]}  x_2^{[\theta]}  
+  \sum_{\theta,\omega} \zeta x_3^{[N+\gamma+\omega]}  x_3^{[\theta]}  
+  \sum_{\theta,\omega} \zeta x_3^{[\omega+N+\gamma]}  x_2^{[\theta]}  
+  \sum_{\theta,\omega} \zeta x_3^{[\omega+N+\gamma]}  x_3^{[\theta]}  
\right)^\lambda \nn\\ 
L_2^{[\gamma,\omega]} & = & \left( x_2^{[\gamma]} x_1^\mu - 
\beta_3 x_3^{[\gamma+\omega]} \right) \left( \ep + 
\alpha x_3^{[\gamma+\omega]} + 
\chi \sum_\theta x_2^{[\theta]} + \chi \sum_\theta x_3^{[\theta]}  
+ \zeta x_3^{[\gamma+\omega]} \sum_\theta x_2^{[\theta]} 
+ \zeta x_3^{[\gamma+\omega]} \sum_\theta  x_3^{[\theta]}   \right)^\mu .
\nn \eeqa
This system has mass density given by 
\beq
\ro = 4 x_1 + 
\sum_{\gamma:|\gamma|=\Lambda} \Lambda x_2^{[\gamma]} + 
\sum_{\gamma:|\gamma|=\Lambda+\mu} (\Lambda+\mu) x_3^{[\gamma]} .
\eeq

Such a complex system is more easily understood if we 
generalise the notation of the previous section.  In the 
special cases where $\mu=\lambda+1$, which we shall 
study in more detail later,  we simplify the notation for short 
chains $x_2^{[\gamma]}$ to $y_n$, where $n$ 
encapsulates the information stored in $\gamma$. Then we can write
\beqa
\dot y_n & =  & \left( x_1^\Lambda - \beta_2 y_n \right) 
\left( \ep + \alpha y_n + \psum \chi y_p + \pqsum \chi y_{p,q} + 
\psum \zeta y_n y_p + \pqsum \zeta y_n y_{p,q} + \right. \lbl{amc-kin-eq1} \\&& 
\left. \hspace*{25mm} + \mpsum \zeta y_{m,n} y_p +
\mpqsum \zeta y_{m,n} y_{pq,} + \mpsum \zeta y_{n,m} y_p + 
\mpqsum \zeta y_{n,m} y_{p,q} \right)^\lambda  \nn \\ && - 
\sum_{r=1}^N \left( y_n x_1^\Lambda - \beta_3 y_{n,r} \right) 
\left( \ep + \alpha y_{n,r} + \chi \psum y_p + \chi \pqsum y_{p,q} + 
\zeta y_{n,r} \psum y_p + \zeta y_{n,r} \pqsum y_{p,q} \right)^\Lambda \nn\\
&& - \sum_{r=1}^N \left( y_n x_1^\Lambda - \beta_3 y_{r,n} \right) 
\left( \ep + \alpha y_{r,n} + \chi \psum y_p + \chi \pqsum y_{p,q} + 
\zeta y_{r,n} \psum y_p + \zeta y_{r,n} \pqsum y_{p,q} \right)^\Lambda 
\nn \\ \hspace*{-3mm} 
\dot y_{m,n} & = & \left( y_m x_1^\Lambda  + y_n x_1^\Lambda - 
2 \beta_3 y_{m,n} \right) \left( \ep + \alpha y_{m,n} + \chi \psum y_p 
+ \chi  \pqsum y_{p,q}  + \zeta y_{m,n} \psum y_p + 
\zeta y_{m,n} \pqsum y_{p,q} \right)^\Lambda . 
\hspace*{-4mm}\nn\\&& \lbl{amc-kin-eq2} \eeqa

This system of ordinary differential equations can be most 
readily understood in terms of the set of macroscopic chemical 
reactions comprising it. Let us use $X_1$ to denote the monomer 
species, $Y_n$ to denote a short chain of type $n$, and 
$Y_{m,n}$ to denote a long chain whose composition is 
identical to that of a short chain of type $m$ joined to a short 
chain of type $n$.  The reactions which the above kinetic 
scheme describes are listed in Table \ref{4tab2}. It will 
be recalled that, owing to the coarse-graining scheme 
applied, the quantities $Y_n$ and $Y_{m,n}$ represent 
``averages'' over sets of chain lengths and sequences.

The constant quoted after each step in the reaction 
mechanism is the rate coefficient associated with 
that process.   We shall not analyse this still 
very complex model here, however, as its complexity 
would be compounded by the subsequent introduction 
of hydrolytic fragmentation processes. Instead,   
the above model will be simplified further, and used 
in the following sections which include hydrolysis 
for the first time.

\begin{table}[bt] 
\[  \begin{array}{rclccl}
\Lambda X_1 & \rightarrow & Y_n && \ep 
        &  \mbox{ uncatalysed formation of short chains} \\ 
Y_n + \Lambda X_1 & \rightarrow & Y_{n,m} && \ep 
        &  \mbox{ uncatalysed formation of long chains} \\  &&&&& \\ 
Y_n + \Lambda X_1 & \rightarrow & 2 Y_n && \alpha 
        & \mbox{ autocatalysis of short chains} \\ 
Y_{m,n} + Y_m + \Lambda X_1 & \rightarrow & 2 Y_{m,n} && \alpha 
        & \mbox{ autocatalysis of long chains} \\  &&&&& \\ 
Y_m + \Lambda X_1 & \rightarrow & Y_m + Y_n && \chi 
        & \mbox{ crosscatalysis of short by short} \\ 
Y_{m,n} + Y_k + \Lambda X_1 & \rightarrow & Y_{m,n} + Y_{k,l} && \chi 
        & \mbox{ crosscatalysis of long by long} \\
Y_{m,n} + \Lambda X_1 & \rightarrow & Y_{m,n} + Y_k && \chi 
        & \mbox{ crosscatalysis of short by long} \\
Y_m + Y_k + \Lambda X_1 & \rightarrow & Y_m + Y_{k,l} && \chi 
        & \mbox{ crosscatalysis of long by short} \\ &&&&& \\ 
Y_p + Y_n + \Lambda X_1 & \rightarrow & 2 Y_n + Y_p && \zeta 
        & \mbox{ ribozymic synthesis of short chains} \\ 
Y_{p,q} + Y_n + \Lambda X_1 & \rightarrow &  2 Y_n + Y_{p,q} && \zeta 
        & \mbox{ ribozymic synthesis of short chains} \\ 
Y_p + Y_{m,n} + \Lambda X_1 & \rightarrow & Y_n  +  Y_{m,n} + Y_p && \zeta 
        & \mbox{ ribozymic synthesis of short chains} \\ 
Y_{p,q} + Y_{m,n} + \Lambda X_1 & \rightarrow & Y_n +  Y_{m,n} + Y_{p,q} && \zeta 
        & \mbox{ ribozymic synthesis of short chains} \\ 
Y_p + Y_{n,m} + \Lambda X_1 & \rightarrow & Y_n + Y_{n,m} + Y_p && \zeta 
        & \mbox{ ribozymic synthesis of short chains} \\ 
Y_{p,q} + Y_{n,m} + \Lambda X_1 & \rightarrow & Y_n +  Y_{n,m}+Y_{p,q} && \zeta 
        & \mbox{ ribozymic synthesis of short chains} \\  &&&&& \\ 
Y_p + Y_{m,n} + Y_m + \Lambda X_1 & \rightarrow  & 2 Y_{m,n} + Y_p && \zeta
        & \mbox{ ribozymic synthesis of long chains} \\
Y_{p,q} + Y_{m,n} + Y_m + \Lambda X_1 & \rightarrow  & 2 Y_{m,n} + Y_{p,q} && \zeta 
        & \mbox{ ribozymic synthesis of long chains} \\
Y_p + Y_{m,n} + Y_n + \Lambda X_1 & \rightarrow  & 2 Y_{m,n} + Y_p && \zeta
        & \mbox{ ribozymic synthesis of long chains} \\
Y_{p,q} + Y_{m,n} + Y_n + \Lambda X_1 & \rightarrow  & 2 Y_{m,n} + Y_{p,q} && \zeta 
        & \mbox{ ribozymic synthesis of long chains} 
\end{array} \]
\caption{{\bf Reactions described in the kinetic equations 
(\protect\ref{amc-kin-eq1})--(\protect\ref{amc-kin-eq2}), 
together with their forward rate constants and a brief description of each.}}
\lbl{4tab2}
\end{table}

\clearpage

To conclude this section, we have shown that a systematic 
coarse-grained contraction procedure can be employed to 
greatly reduce the number of rate equations and different 
intermediates that need to be considered.  Such a 
procedure yields equations which are amenable 
to theoretical as well as numerical analysis and we 
shall show in Section~\ref{det-hydrolysis} 
how the most drastic approximation--the maximally 
contracted system--can still account for the 
way in which a combination of template-based synthesis 
and ribozymic replicase activity amplifies initial 
differences in RNA chain concentrations. 
In Section~\ref{red-hydrolysis}, we shall return to 
the `almost maximally contracted 
system' when incorporating hydrolysis into the
theoretical description of RNA chain replication. 

\section{RNA formation including hydrolysis} \lbl{det-hydrolysis}
\setcounter{equation}{0}

{}From a chemical point of view, a 
crucial omission from all of the foregoing models is hydrolysis. 
This is the process by which polymeric ribonucleotide 
chains are prevented from growing 
to unlimited lengths.  Many of the models analysed above 
have the unfortunate property of producing 
infinitely long chains in a finite time.  This implies the presence of
mathematical singularities within the kinetic equations and by itself 
indicates that certain essential physicochemical 
processes have been overlooked. 

So far, our general models of RNA formation have been based on the
\BD\ theory, whereby chains grow or fragment through {\em one step}
addition or removal of $\beta$-D-ribonucleotide monomers. For more
general coagulation-fragmentation processes, in which arbitrarily
large amounts of polymer may be added or removed, one can use a
description based on the general Smoluchowski equations~(\ref{Smol})  
instead of the \BD\ equations. While the \BD\ theory evidently works 
well for the stepwise growth of RNA chains, hydrolysis can only be 
accounted for by a general fragmentation term, since the loss of a 
single monomer from the end of an RNA oligomer is an exceedingly poor 
approximation to hydrolysis.  Therefore, in the models described 
below we shall retain the original \BD\ assumptions when modelling 
the aggregation mechanism, but we shall include a general Smoluchowski 
fragmentation term to handle the effects of hydrolysis.  

We wish to emphasize that we do not make any particular assumptions
about the nature of hydrolysis. We certainly do not seek to describe
it in terms of equilibrium proceses (which would enable an arbitrary
hydrolytic chain fragmentation to be equivalently described in terms
of a sequence of stepwise processes, and to which Le Chatelier's
principle could be applied). Rather, hydrolysis is modelled explicitly
by a distinct kinetic process which splits a chain in two at an 
arbitrary point and is thus of general Smoluchowski fragmentation form.

In the rest of this section, our model of possibly catalytic \BD\ chain 
growth combined with a general fragmentation term will be analysed by 
the use of generating function and coarse-grained reduction techniques. 

\subsection{Formulation of the Smoluchowski
coagulation-fragmentation equations for RNA formation with hydrolysis}

The full Smoluchowski coagulation-fragmentation equations for the rate
of change of concentration of a polymeric chain of length $r$ are
\beq \begin{array}{rclrl}
\dot{c}_r & = & 
\half \ds\sum_{k=1}^{r-1} W_{k,r-k} - \ds\sum_{k=1}^\infty W_{r,k} , 
&\;\;\;\;\;\;\;& r=2,3,\ldots\\
W_{r,s} & = & a_{r,s} c_r c_s - b_{r,s} c_{r+s} , &&  \lbl{Smol}
\end{array} \eeq
where the rate coefficients for the aggregation and fragmentation
steps $a_{r,s},b_{r,s}$ respectively now carry two suffices,
indicative of the fact that clusters of sizes $r$ and $s$ can
reversibly coalesce to form a cluster of size $(r+s)$.
We shall analyse two versions of this system: in one, the 
monomer concentration $c_1$ will be assumed 
constant; in the other, the monomer concentration will be 
allowed to vary, and the density $\ro = \sum_{r=1}^\infty r c_r$ 
will be held constant. 

To recover the original \BD\ equations from (\ref{Smol}), 
we impose the condition that there can be no polymer-polymer 
aggregation, so that a ribonucleotide chain can only 
aggregate {\em via} one monomer at a time.  
Thus we impose the following form 
\beq
a_{r,s} = a_r \delta_{s,1} + a_s \delta_{r,1} , 
\eeq
on the forward coefficients $a_{r,s}$ (note that this implies
$a_{11}=2a_1$); where, $\delta_{i,j}$ is the Kronecker delta 
satisfying $\delta_{i,j}=1$ only if $i=j$ and is zero otherwise. 

Just as template-based catalysis was included in the \BD\ model of Section 
\ref{detail-model-sec}, it can be incorporated within this more 
general coagulation--fragmentation model.  This is achieved 
by replacing $a_{r,s}$ above by $a_{r,s} + \ksum f_{r,s,k} c_k$, where
$f_{r,s,k}$ is another set of rate coefficients for the template
synthesis of oligomeric $\beta$-D-ribonucleotides. 

We shall make extensive use of generating function techniques, 
and choose simple forms for the coefficients $a_{r,s},b_{r,s}$ 
so that these techniques may be used to their full potential. 
Initially, in the same manner as in equation (\ref{Fdef}), 
we define the generating function 
\beq
C(z,t) = \ds\rsum c_r(t) e^{-zr}  , \lbl{genfndef}
\eeq
which can be viewed as a discrete Laplace transform. 
This transforms the problem from having one discrete 
and one continuous independent variable $(r,t)$ to two 
continuous variables $(z,t)$. Note that putting $z=0$ 
in the above expression gives the quantity $C_0(t) = C(0,t)$ 
which is the total number  of chains in the system.  
A second important quantity in the analysis of these 
problems is 
\beq
u(z,t) = - \pad{C}{z} = \rsum r c_r e^{-zr} , 
\eeq
and inserting $z=0$ in this yields the density, $\ro=u_0(t) = u(0,t)$.

In the main, we shall choose constant rate coefficients, although
forward rate coefficients of the form $a_r = a + \tilde{a} r$ can be 
analysed without much extra difficulty.  Two types of 
backward rate coefficients can also be analysed: (i) $b_{r,s}=b$, which 
we shall concentrate on, and (ii) $b_{r,s} = b/(r+s-1)$ which we 
do not consider further here.  The choice of a constant hydrolysis 
rate implied by (i) is a reasonable 
first approximation; the choice (ii), where long chains have a 
smaller chance of undergoing hydrolysis than short ones, is less tenable. 

As was done earlier in the paper, two types of 
prebiotic scenario are analysed below. The first has 
constant mass density $\rho$, for which a simple closed system of 
equations can be generated for the monomer concentration ($c_1(t)$) 
and the total concentration of polymer chains including monomers 
($C_0(t)$).  These give some indication 
of the polydispersity of the chains present in the soup since the 
average RNA chain length is then equal to 
\beq
{\bf E}(t) = \langle r \rangle = \frac{ \sum r c_r(t) } { \sum c_r(t) } =
\frac{\ro}{C_0(t)} . 
\eeq
Typically we are interested in the evolution of the system 
starting from an initial condition where all the mass is in 
monomeric form; that is $c_1(0)=\ro$, $c_r(0)=0\;\forall r\geq2$ 
which implies $C_0(0)=\ro$.  Systems of this type are analysed 
in Sections \ref{det-hy-c1-sec1} and \ref{det-hy-c1-sec2}. 

In Section \ref{det-hy-ro-sec}, a second prebiotic scenario is 
analysed, which has a fixed monomer concentration ($c_1$), so 
that the total ribonucleotide density $u_0(t)$ may vary. This is
equivalent to the
pool chemical approximation. 
In this case we can form a simple closed system of equations for 
the density $u_0(t)$ and the total number of polymer chains $C_0(t)$.

In all situations we shall initially consider, the only type 
of template-based synthesis we analyse involves 
error-prone replication or crosscatalysis  (in fact, pure 
autocatalysis cannot be analysed using generating 
function methods as the type of nonlinearity which it 
introduces is not amenable to transform techniques).   
For this reason we have not given the full formulation of 
the problem, including details of individual sequences, 
as that would introduce autocatalysis both directly 
and as part of the ribozymic replicase polymerization 
mechanism.  However, these more complex mechanisms will be 
introduced in the next section (Section \ref{red-hy-sec}) 
when we consider massively contracted models. 

\subsection{Constant monomer concentrations; pure template-based RNA synthesis}
\lbl{det-hy-c1-sec1}

In this section we shall reconsider the case analysed in the Appendix 
which has forward rate constants proportional to chain length; there, 
the total mass in the system and the average chain length both blew up in 
finite time. We now show that the incorporation of hydrolysis into the 
model prevents such unrealistic behaviour.   

Whereas an RNA chain can grow only by adding a single monomer to 
one or other end of the chain, a chain can split into two parts at any 
position along its length.    Since the general fragmentation term 
we have included gives each phosphate bond an equal chance of breaking, 
and the number of bonds in a chain is proportional to chain length, 
the reverse reaction rate coefficient is effectively proportional to 
chain length.  Thus we expect that a general hydrolytic fragmentation 
mechanism with constant rate coefficients will be able to prevent
unbounded chain growth in a polymerization mechanism 
with forward rate constants proportional to chain length.   

The equations which describe pure template synthesis 
together with hydrolysis 
in a system with constant ribonucleotide monomer concentration are 
\beq
\dot c_r = \ssum f (r-1) s c_s c_{r-1} c_1 - \ssum f r s c_s c_r c_1 - 
\half b (r-1) c_r + \ssum b c_{r+s} . 
\eeq

The generating function $C(z,t)$ of equation~(\ref{genfndef}) 
is then determined by 
\beq
\pad{C}{t} = f c_1^2 u_0 e^{-z} - f c_1 u_0 u (1-e^{-z}) 
- \half b u + \half b C + b c_1 e^{-z} + 
b\left( \frac{ C_0 e^{-2z} - C}{1-e^{-z}}\right) , \lbl{det-hy-pcat-Ceq}
\eeq
where $u=-\pad{C}{z}$ in turn satisfies 
\beqa
\pad{u}{t} & = & f c_1^2 u_0 e^{-z} + f c_1 u_0 \pad{u}{z} (1-e^{-z}) + 
f c_1 u_0 u e^{-z} + \half b \pad{u}{z} + \half b u + b c_1 e^{-z} \nn \\&&
-bC_0e^{-z}-b\left(\frac{u(1-e^{-z})+(C-C_0)e^{-z}}{(1-e^{-z})^2}\right).
\lbl{det-hy-pcat-ueq} \eeqa
A closed system can then be formed for the total number of 
chains $C_0(t)$ and total mass of RNA chains $u_0(t)$ by letting 
$z\rightarrow0$ in (\ref{det-hy-pcat-Ceq}) and
(\ref{det-hy-pcat-ueq}), giving
\beqa
\dot C_0 & = & f c_1^2 u_0 + \half b u_0 - \mfrac{3}{2} b C_0 + b c_1 \\
\dot u_0 & = & f c_1 u_0^2 + f c_1^2 u_0 - b C_0 + b c_1 . 
\eeqa
In general, explicit solutions of this pair of equations 
are not available, but special cases can be analysed, 
and a fuller picture built up from these. 

The system has equilibrium solutions at 
\beq
\ol{u}_0=\frac{c_1}{6} \left( \frac{b}{fc_1^2} - 1 \pm 
\sqrt{ \frac{b^2}{f^2c_1^4} - \frac{14 b }{fc_1^2} + 1 } \right) , 
\eeq
where this is well-defined, namely for $b/fc_1^2\geq 7+\sqrt{48}\approx 13.9$ 
(for $b/fc_1^2\leq7-\sqrt{48}\approx0.07$, both roots correspond 
to negative values of the density $\ol{u}_0$).

\subsubsection{The case {$b=0$}}

For $b=0$ the system of two ordinary differential 
equations is solvable.  Making the substitution 
$u_0 = 1/w$ produces a linear equation which can 
be solved, and leads to 
\beq
u_0(t) = \frac{c_1}{2 e^{-fc_1^2 t} -1} ; 
\eeq
thus the system blows up at time $t=t_g=\log(2)/fc_1^2$ 
which, as is to be expected, is the result obtained 
in the Appendix, since $b=0$ corresponds to no hydrolysis.  
For small $b$ there will be only minor modifications to 
this behaviour.  

\subsubsection{The case {$b=b_c=(7+4\surd{3})fc_1^2$}}

At the smallest value of $b$ where an equilibrium solution exists, 
the equilibrium values of $C_0$ and $u_0$ are $\rec{3}(1\!+\!2\sqrt{3})c_1$
and $(1\!+\!2/\sqrt{3})c_1$ respectively. An expansion about the 
equilibrium point using 
\beqa
C_0 = \rec{3} c_1 (1+2\sqrt{3}) (1+g) , \hspace{5mm} 
u_0 = c_1 (1+\mfrac{2}{3}\sqrt{3})(1+w) 
\eeqa
reveals the existence of one stable and one centre manifold~\cite{Carr}.
The existence of a centre manifold is to be expected since at $b=b_c$ 
the system undergoes a saddle-node bifurcation:  
\beqa
\frac{1}{f c_1^2} \frac{d g}{d t} & = & 
\left( \frac{3}{10\sqrt{3}-17} \right) \left( 
- (\sqrt{3}+\half) g + (\sqrt{3}-\half) w \right) , \\
\frac{1}{f c_1^2} \frac{d w}{d t} & = & 
\left( \frac{2}{2\sqrt{3}-3} \right) \left( 
- (\sqrt{3}+\half) g + (\sqrt{3}-\half) w + \mfrac{3}{4} w^2 \right) , 
\eeqa
The eigenvector corresponding to the stable manifold is 
$\left({{21/2+6\surd3}\atop{5+8/\surd3}}\right)$ and that 
corresponding to the centre manifold is 
$\left({{\surd3-1/2}\atop{\surd3+1/2}}\right)$.  
This situation is semi-stable in that the equilibrium 
position is only stable from one side.  If too much 
matter is inserted into the system, then the system's 
behaviour is divergent, gaining mass without bound. 
However, if the system is initiated with only a little 
matter present, it will evolve smoothly and slowly approach 
an equilibrium solution.  The situation is summarised 
in Figure 1. 

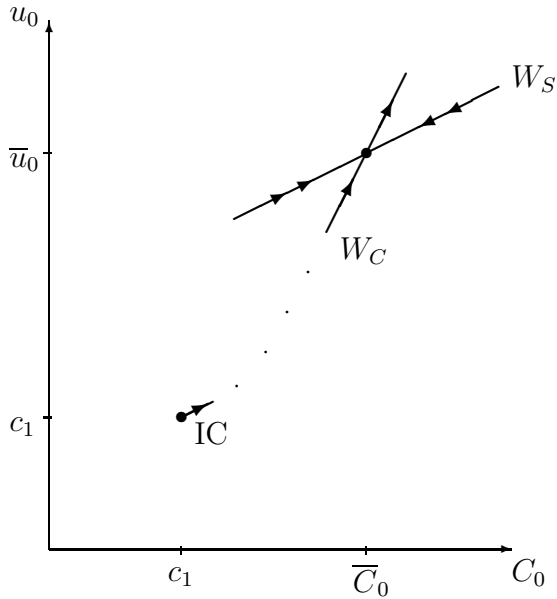
\begin{figure}[hbt]
\vspace{2.5in}
\begin{picture}(300,30)(-60,-10)
\thinlines
\put(00,00){\vector(1,0){175}}
\put(00,00){\vector(0,1){200}}
\put(50,50){\circle*{4}}
\put(50,-2){\line(0,1){3}}
\put(-2,50){\line(1,0){3}}
\put(120,-2){\line(0,1){3}}
\put(-2,150){\line(1,0){3}}
\put(45,-12){$c_1$}
\put(-15,45){$c_1$}
\put(175,-12){$C_0$}
\put(-15,200){$u_0$}
\put(115,-16){$\ol{C}_0$}
\put(-15,145){$\ol{u}_0$}
\put(55,40){IC}
\thicklines
\put(50,50){\vector(2,1){10}}
\put(52,51){\line(2,1){10}}
\put(120,150){\circle*{4}}
\put( 70,125){\vector(2,1){20}}
\put( 90,135){\vector(2,1){10}}
\put(100,140){\line(2,1){40}}
\put(170,175){\vector(-2,-1){20}}
\put(150,165){\vector(-2,-1){10}}
\put(175,175){$W_S$}
\put(105,120){\vector(1,2){10}}
\put(110,130){\vector(1,2){20}}
\put(135,180){\line(-1,-2){10}}
\put(110,110){$W_C$}
\put(71,62){\circle*{1}}
\put(82,75){\circle*{1}}
\put(90,90){\circle*{1}}
\put(98,105){\circle*{1}}
\end{picture}
\caption{{\bf Diagram of the ($C_0,u_0$) phase plane 
for the case $b=b_c$. IC marks the initial condition 
imposed on the system where all nucleotide matter is 
in monomeric form and so both the mass $u_0=c_1$, and 
number of objects in the system $C_0=c_1$. $W_C$ is the 
centre manifold emanating from the equilibrium configuration 
of the system, along which the system slowly moves.  $W_S$ 
marks the stable manifold of the equilibrium point; its 
existence shows that the centre manifold is itself stable 
in that points close to the stable manifold are attracted 
to it. At equilibrium the number of chains $C_0\!=\!\ol{C}_0$ 
and the mass $u_0\!=\!\ol{u}_0$.}}
\label{fig2}
\end{figure}

Systems which start below the stable manifold -- 
the lines marked $W_S$ on Figure \ref{fig2} -- are 
attracted to the equilibrium solution along the 
centre manifold ($W_C$); however, systems 
initiated above $W_S$ gain mass and are 
repelled from the equilibrium point. Thus the 
trajectory we are interested in (that starting 
from the initial conditions $C(0)=c_1-u_0(0)$) 
does approach the equilibrium solution. 

\subsubsection{The case $b\gg1$}

For asymptotically large $b$, the system has a critical point near 
\beq
C_0 \sim c_1 \left( 1 + \frac{2fc_1^2}{b} \right) , \hspace{7mm} 
u_0 \sim c_1 \left( 1 + \frac{4fc_1^2}{b} \right) .
\eeq
The nature of this critical point is a stable node,  
so the initial conditions at $C_0=u_0=c_1$ are 
attracted to this stable equilibrium configuration. 
(There is also a saddle point at $u_0 \sim b/3fc_1$, 
$C_0 \sim b/9fc_1$.) This is intuitively in line with 
what we expect--$b$ very large implies that hydrolysis 
dominates the kinetics, making it almost impossible 
to build long chains; the few that are built rapidly 
break up, so that the system remains mainly in monomeric form. 

\subsubsection{Summary} 

This example shows that the inclusion of hydrolysis, according to which
RNA chains split at any point along their length, provides a powerful 
mechanism which can prevent chains from growing infinitely long.   
With a constant hydrolysis rate (independent of chain length) the 
mechanism prevents infinitely long chains forming in RNA systems in 
which the growth rate is proportional to chain length.  

In Section \ref{cross-sec}, a model with coefficients 
independent of chain length was analysed and shown 
to have divergent behaviour in the limit $t\rightarrow\infty$. 
In this limit, both the number of chains and mass of material 
in chains grow without bound in such a way that the 
expected chain length also grows unboundedly.  
In the Appendix it is shown that in models where 
rate constants are proportional to chain length, 
such a singularity can occur in finite time.  

This section has shown that if a significant degree of 
hydrolysis is included into such models, it prevents 
this unphysical behaviour.  If the hydrolysis rate 
exceeds a threshold value,  there is a stable equilibrium 
configuration to which the system will tend. 

\subsection{Constant rate coefficients and constant nucleotide 
monomer concentrations}
\lbl{det-hy-c1-sec2}

We now consider a more general model which includes 
both catalysed and uncatalysed RNA chain growth mechanisms. 
We assume that all reaction rates are independent of 
chain length, using $f$ to denote the catalysed rate and 
$a$ the uncatalysed rate of addition of nucleotide 
monomers to a chain. 

For $r>1$, the equation 
\beq
\dot{c}_r = a c_1 c_{r-1} - a c_1 c_r + f c_1 c_{r-1} \ssum c_s -
f c_1 c_r \ssum c_s - \half b (r-1) c_r + b \ssum c_{r+s}  , 
\eeq
governs the rate of change of concentrations. The generating 
function then satisfies 
\beqa
\pad{C}{t} & = & ac_1^2e^{-z}+fc_1^2C_0e^{-z}-ac_1C(1-e^{-z}) 
-fc_1C_0C(1-e^{-z}) + \nn\\&& + \half bC - \half b u + b c_1 e^{-z} 
+ \frac{b(C_0 e^{-2z}-C)}{(1-e^{-z})} , 
\eeqa
where $u=-\pad{C}{t}$ satisfies the partial differential equation 
\beqa
\pad{u}{t} & = & c_1 e^{-z} (c_1+C)(a+fC_0)- 
u c_1(1-e^{-z})(a+fC_0) + \nn\\&& + \half b u + \half b \pad{u}{z} + 
b c_1 e^{-z} - b C_0 e^{-z}  -  
\frac{b[(1-e^{-z})u + e^{-z}(C-C_0)]}{(1-e^{-z})^2} . 
\eeqa
Again, these equations cannot be solved in general, 
but useful information about their solutions can be 
obtained by considering the limit $z\rightarrow0$ 
in which they reduce to a coupled pair of ordinary 
differential equations: 
\beqa
\dot C_0 & = & \left( f c_1^2 - \mfrac{3}{2} b \right) C_0 + 
a c_1^2 + b c_1 + \half b u_0 \nn \\
\dot u_0 & = & f c_1C_0^2+( f c_1^2+ac_1-b)C_0+ac_1^2+bc_1 .
\lbl{83kineq} \eeqa

We seek an equilibrium solution of these equations by first solving 
the latter quadratic equation to find $C_0$ and then the former linear 
expression which gives $u_0$ in terms of $C_0$. 
Let us define 
\beq
B = \frac{b}{ac_1+fc_1^2} , \hspace{9mm} 
\chi = \frac{fc_1^2}{ac_1+fc_1^2} , 
\eeq
so that $B$ is the reciprocal of the equilibrium constant for the
overall reaction (the ratio of the total hydrolytic fragmentation 
rate to the total polymerization rate), while $\chi$ 
is a measure of the effectiveness of catalysis, since it is the 
ratio of the catalysis rate to the total aggregation rate ($0<\chi<1$).  
Thus the equilibrium value of $C_0$ is given by 
\beq
\ol{C}_0=\frac{1}{2\chi} \left(B-1\pm\sqrt{ (B-1)^2-4\chi }\right).
\eeq
Due to the form of the discriminant, if the roots for $\ol{C}_0$ are real 
then they have the same sign.  If $B<1$ then they are both negative 
or both complex and hence  are not relevant for the large time 
asymptotics of the system (\ref{83kineq}). In this case the total 
number of chains and mass of material must both tend to infinity.    
If $1<B<1+2\sqrt{\chi}$, then the roots for $\ol{C}_0$ are complex 
and hence again irrelevant when considering the large time 
asymptotics of the system (\ref{83kineq}).  However, if 
$B>1+2\sqrt{\chi}$ then hydrolysis is strong enough 
to prevent such unphysicochemical divergences, and in the large-time limit the 
system may be attracted to such an equilibrium configuration. 

\subsection{Constant rate coefficients and constant nucleotide density}
\lbl{det-hy-ro-sec}

In order to consider the constant-density form of the equations, 
we need to construct a new equation for the monomer concentration 
which accounts for the loss of monomer as the ribonucleotide chains grow.
With constant reaction rate coefficients, the kinetic equations are 
\beqa
\dot{c}_r & = & a c_1 c_{r-1} - a c_1 c_r + f c_1 c_{r-1} \ssum c_s -
f c_1 c_r \ssum c_s - \half b (r-1) c_r + b \ssum c_{r+s}  \\
\dot{c}_1 & = & - a c_1^2 - a c_1 \ssum c_s - f c_1^2 \ssum c_s - 
f c_1 \left(\ssum c_s\right) \left(\ssum c_s\right) + b \ssum
c_{s+1}, 
\eeqa
which can again be analysed using generating function techniques; 
from (\ref{genfndef})
\beq
\pad{C}{t} = c_1 e^{-z} ( C - C_0 )(a+f C_0) - c_1 C(a+fC_0) + 
b e^{-z} C_0 + \half b C - \half b u + \frac{b(C_0e^{-2z}-C)}{(1-e^{-z})} .
\lbl{84gfeq} \eeq
In this case the equation for $\pad{u}{t}$ is not so important for us 
since the imposition of constant density implies $\dot{u}_0=0$. 
This leads to the following coupled pair of ordinary differential equations
for the total number of chains and the monomer concentration 
\beqa
\dot C_0 & = & - f c_1 C_0^2 - (a c_1 + \half b) C_0 + \half b\ro \\
\dot c_1 & = & -(a+f C_0)c_1^2 - (aC_0+ f C_0^2+b) c_1 + bC_0.
\eeqa
Starting from initial conditions in which all the available
ribonucleotide mass is in 
monomeric form, which implies $c_1(0)=\ro$, $C_0(0)=\ro$, 
we expect the system to approach an equilibrium configuration with 
significantly less monomers, and fewer chains, but where 
the average chain length $\langle r \rangle := \ro / C_0$ 
is much larger than unity.  

It is not possible to explicitly find equilibrium points of this system, 
but we can show that they exist and occur in the correct 
region of phase space.  First note that $\dot C_0=0$ implies 
that at equilibrium 
\beq
\ol{c}_1 = \frac{ b (\ro-\ol{C}_0) }{ 2 \ol{C}_0 (a+f\ol{C}_0) }. \lbl{24c1bar}
\eeq
Then the condition $\dot c_1=0$ corresponds to finding roots of the function 
\beq
G(\ol{C}_0) := b (\ro-\ol{C}_0)^2 + 2 \ol{C}_0 (\ro-\ol{C}_0) 
(f \ol{C}_0^2 + a \ol{C}_0 + b) - 4 \ol{C}_0^3 ( a + f \ol{C}_0 ) .
\eeq
{}From this it is possible to see that $G(0)=b^2\ro>0$ and
$G(\ro)=-4b\ro^3(a+f\ro)<0$ so there is an equilibrium  value 
of $\ol{C}_0$ between zero and $\ro$; and from (\ref{24c1bar}) 
we can find a corresponding value of $\ol{c}_1$.

In the limit $b\rightarrow0$, it is possible to obtain the equilibrium
solution asymptotically; in this case it is 
\beq
\ol{C}_0=\rec{3}\ro , \hspace{6mm}  
\ol{c}_1=3b/(3a+f\ro) , 
\eeq
so the monomer concentration is small and the equilibrium distribution 
of chains has mean length $\langle r\rangle=\ro/\ol{C}_0=3$. 

The ribonucleotide polymer chain 
distribution can then be found by solving (\ref{84gfeq}) with 
$\pad{C}{t}=0$ and $u=-\pad{C}{z}$ to give 
\beq
\ol{C}(z) = \rec{3} \ro \left[ 1 - \mfrac{3}{2} e^z(e^{2z}-1) +
\mfrac{3}{2} e^{z+2e^{-z}} (e^z-1)^2 \right] , 
\eeq
from which we can find the concentration distribution 
\beq
\ol{c}_r = \frac{\ro \, 2^r \, (r\!-\!1) }{ (r\!+\!1)! \; (r\!+\!3) } . 
\eeq
This corresponds to a pure aggregation \BD\ model 
where coagulation stops as $c_1\rightarrow0$.  Figure 
\ref{84fig} shows this profile which demonstrates 
how effective even a small amount of hydrolysis 
can be in moderating the growth of long chains. 

\begin{figure}[hbt]
  \vspace{2.5in}
  \includegraphics{fig84c.ps}
  \includegraphics{fig84logc.ps}
  \caption{{\bf Graphs of (a) concentration and (b) log (concentration) 
  against ribonucleotide chain length ($r$) for the case with constant 
  rate coefficients and constant total nucleotide density in the limit 
  $b\rightarrow0$, i.e. for vanishingly small hydrolysis (so that $\ol{c}_1=0$).}}
  \label{84fig}
\end{figure}

\section{Coarse-grained models of RNA formation including hydrolysis}
\setcounter{equation}{0} \lbl{red-hy-sec}
\lbl{red-hydrolysis}

The most straightforward way to incorporate hydrolysis into 
the contracted RNA models is to modify the ``almost maximally 
contracted'' model from Section \ref{amc-subsec}, by 
adding extra terms to the right-hand side of
(\ref{amc-kin-eq1})--(\ref{amc-kin-eq2}). 
By the term ``almost maximally contracted'' we mean 
a coarse-grained reduction which describes monomers 
and two types of chain: short and long.  (Recall that we 
imply here {\em sets} of ribonucleotide sequences, not 
individual RNA polymer molecules.)   Then we can add in one 
extra mechanism--namely hydrolysis--which replaces 
one long chain by two short ones.  However, as noted 
previously, since hydrolysis violates the \BD\ assumptions 
that proscribe cluster--cluster interactions, we cannot 
{\em a priori} build hydrolysis into a full \BD\ model, and 
then perform the coarse-graining contraction.  Instead we take 
an already contracted model and add hydrolysis {\em a 
posteriori}. We return to the almost maximally contracted model 
described by the rate processes listed in Table \ref{4tab2},  
and denote  short chains (of length $\Lambda$) by $Y_n$, 
where the subscript $n$ enumerates the variety of different 
possible compositions of chains (ribonucleotide sequences), 
and long chains (length $2\Lambda$) by $Y_{m,n}$.  
Here as before $m$ represents the information carried 
in the first half of the chain and $n$ the information 
in the second half.  Thus the chain denoted $Y_{m,n}$ 
is simply a concatenation of the chain $Y_m$ with $Y_n$.  

We shall not analyse the full system of rate processes 
described in Table \ref{4tab2} as, using physicochemical 
insight, one can see immediately that some 
of the steps quoted there make a negligible contribution 
to the replication of RNA.  The mechanisms which we aim to 
model are described in Table \ref{6tab-inc}.   
The constant quoted after each reaction mechanism is the 
rate coefficient associated with that process. 
A number of possible mechanisms have been eliminated from our 
model; these are listed in Table \ref{6tab-reject}. 

We assume that chains are subject to hydrolysis (which 
splits one long chain into two short ones), and 
that both types of chain (short and long) are grown by an uncatalysed 
mechanism as well as autocatalytically.  We shall 
assume that chains can only act as crosscatalysts 
to chains of the same length or shorter; thus 
we neglect the putative mechanism by which long chains 
are formed with a short chain acting as a catalyst 
as the rate constant for this interaction is expected 
to be very small. 
This assumption is carried through to the ribozymic 
synthesis mechanisms where we assume that only 
longer chains can act as enzymatic catalysts. 

\begin{table}[tb] 
\[ \hspace*{-5mm} \begin{array}{rclccl}
\Lambda X_1 & \rightarrow & Y_n && \epS 
        &  \mbox{ uncatalysed polymerisation} \\ 
Y_n + \Lambda X_1 & \rightarrow & Y_{n,m} && \epL 
        &  \mbox{ uncatalysed growth of long from short chains} \\  
Y_n + \Lambda X_1 & \rightarrow & 2 Y_n && \alphaS 
        & \mbox{ autocatalysis of short chains} \\ 
Y_{m,n} + Y_m + \Lambda X_1 & \rightarrow & 2 Y_{m,n} && \alphaL 
        & \mbox{ autocatalysis of long chains} \\  
Y_{m,n} & \rightarrow & Y_m + Y_n  &&  \eta & \mbox{ hydrolysis } \\ 
                                                &&&&& \\ 
Y_m + \Lambda X_1 & \rightarrow & Y_m + Y_n && \chiS 
        & \mbox{ crosscatalysis of short by short chains} \\ 
Y_{m,n} + Y_k + \Lambda X_1 & \rightarrow & Y_{m,n} + Y_{k,l} && \chiL 
        & \mbox{ crosscatalysis of long by long chains} \\
Y_{m,n} + \Lambda X_1 & \rightarrow & Y_{m,n} + Y_k && \chiX 
        & \mbox{ crosscatalysis of short by long chains} \\
                                                &&&&& \\ 
Y_{p,q} + Y_n + \Lambda X_1 & \rightarrow &  2 Y_n + Y_{p,q} && \zetaSS 
        & \mbox{ ribozymic synthesis of short chains} \\ 
Y_{p,q} + Y_{m,n} + \Lambda X_1 & \rightarrow & Y_n +  Y_{m,n} + Y_{p,q}&&\zetaSL
        & \mbox{ ribozymic synthesis of short chains} \\ 
Y_{p,q} + Y_{n,m} + \Lambda X_1 & \rightarrow & Y_n +  Y_{n,m}+Y_{p,q} &&\zetaSL
        & \mbox{ ribozymic synthesis of short chains} \\  
Y_{p,q} + Y_{m,n} + Y_m + \Lambda X_1 & \rightarrow  & 2 Y_{m,n}+Y_{p,q}&&\zetaLL
        & \mbox{ ribozymic synthesis of long chains} \\
Y_{p,q} + Y_{m,n} + Y_n + \Lambda X_1 & \rightarrow  & 2Y_{m,n}+Y_{p,q}&&\zetaLL
        & \mbox{ ribozymic synthesis of long chains} 
\end{array} \]
\caption{{\bf Reactions included in the almost maximally contracted 
model (\protect\ref{shortdot})--(\protect\ref{longdot}), together with 
their corresponding forward rate constants. Subscripts `$S$',`$L$'  denote 
short and long chains respectively.}}
\lbl{6tab-inc} \end{table}

\begin{table}[tb] 
\[ \begin{array}{rclccl}
Y_m + Y_k + \Lambda X_1 & \rightarrow & Y_m + Y_{k,l} && \chi 
        & \mbox{ crosscatalysis of long by short chains} \\ 
Y_p + Y_n + \Lambda X_1 & \rightarrow & 2 Y_n + Y_p && \zeta 
        & \mbox{ ribozymic synthesis of short chains} \\ 
Y_p + Y_{m,n} + \Lambda X_1 & \rightarrow & Y_n  +  Y_{m,n} + Y_p && \zeta 
        & \mbox{ ribozymic synthesis of short chains} \\ 
Y_p + Y_{n,m} + \Lambda X_1 & \rightarrow & Y_n + Y_{n,m} + Y_p && \zeta 
        & \mbox{ ribozymic synthesis of short chains} \\ 
Y_p + Y_{m,n} + Y_m + \Lambda X_1 & \rightarrow  & 2 Y_{m,n} + Y_p && \zeta
        & \mbox{ ribozymic synthesis of long chains} \\
Y_p + Y_{m,n} + Y_n + \Lambda X_1 & \rightarrow  & 2 Y_{m,n} + Y_p && \zeta
        & \mbox{ ribozymic synthesis of long chains} 
\end{array} \]
\caption{{\bf Reactions not included in the model equations 
(\protect\ref{shortdot})--(\protect\ref{longdot}).}}
\lbl{6tab-reject}
\end{table}

Another assumption made in the model is that 
the main effect of hydrolysis is to split long 
chains into short ones, and not short chains 
into monomers. 
The reaction mechanism $Y_n\rightarrow \Lambda X_1$ could be 
added to the scheme but would further complicate the equations. 

The kinetic equations for the reactions listed in Table \ref{6tab-inc} are  
similar to (\ref{amc-kin-eq1})--(\ref{amc-kin-eq2}); although slightly 
simpler due to the elimination of some mechanisms, they have 
one extra term introduced by the inclusion of hydrolysis 
\beqa \hspace*{-6mm} 
\dot y_n \!& \!=\! & \!\msum \eta ( y_{m,n} \!+\! y_{n,m} )  -
\msum \left( x_1^\Lambda y_n \!-\! \beta_3 y_{m,n} \right) \left( 
\epL \!+\! \alphaL y_{m,n} \!+\! \pqsum \chiL y_{p,q} \!+\! 
\pqsum \zetaLL y_{m,n}y_{p,q}  \right)^\Lambda \nn\\ && 
- \msum \left( x_1^\Lambda y_n \!-\! \beta_3 y_{n,m} \right) \left( 
\epL \!+\! \alphaL y_{n,m} \!+\! \pqsum \chiL y_{p,q} \!+\! 
\pqsum \zetaLL y_{n,m}y_{p,q}  \right)^\Lambda \lbl{shortdot}\\&& 
\!\!\!\!\!\!\!\! + \left( x_1^\Lambda \!-\! \beta_2 y_n \right) \left( 
\epS \!+\! \alphaS y_n \!+\! \msum \chiS y_m \!+\! 
\pqsum \chiX y_{p,q}\!+\! \pqsum \zetaSS y_{p,q} y_n \!+\! 
\mpqsum \zetaSL y_{p,q} (y_{m,n}\!+\!y_{n,m}) \right)^\lambda  \nn\\
\!\!\!\! \dot y_{m,n}\! & \!=\! & \!- \eta y_{m,n} + \left( 
x_1^\Lambda y_m \!+\! x_1^\Lambda y_n \!-\! 2 \beta_3 y_{m,n} 
\right) \left(  \epL \!+\! \alphaL y_{m,n} \!+\! \pqsum \chiL y_{p,q} 
\!+\! \pqsum \zetaLL y_{m,n}y_{p,q} \right)^\Lambda . \lbl{longdot}
\hspace*{-3mm} \eeqa

As a first approximation all the reactions are taken to be 
irreversible.  Another way of interpreting this is to say that 
the dominant chain-shortening mechanism is hydrolysis;  
to leading order we ignore the reversibility of the stepwise 
growth mechanisms (this corresponds to setting $\beta_2 = 0 = \beta_3$).

Initially we will analyse the model with constant ribonucleotide monomer 
concentrations, mainly because this further simplifies the system, 
but later we examine the constant density system.  
The large number and complicated nature of the nonlinear terms
present in this system of equations mean that a general solution
cannot be found from analytic techniques.  
A full solution is only available using numerical methods; 
however, even this is fraught with difficulty due to the 
combinatorially large number of equations in the system.  
Here we shall study special solutions analytically in order 
to determine certain important generic features of the kinetics 
allowed by these equations.  As in the analysis presented 
in Section \ref{mc-thry-sec}, we shall look for uniform 
solutions in which all 
concentrations of short chains are equal, and all concentrations 
of long chains are equal.  Once certain properties 
of such a solution have been established, we prove that 
these uniform solutions are unstable to perturbations which cause some 
types of chain to grow at the expense of others. 

\subsection{Constant ribonucleotide monomer concentration}
\label{61sec}

If the monomer concentration is taken as constant (pool chemical
approximation) then 
the system of rate equations is slightly simplified.  For simplicity we shall 
scale the concentrations so that $x_1=1$.  As noted above, these 
assumptions imply that $\beta_2=0=\beta_3$.  

\subsubsection{Uniform solution}

To find this simple uniform solution, for which the concentrations of all
short chains are equal, and similarly for the long chains, 
we assume that all the short chains $y_n(t)$ 
have the same concentration (say $Y(t)$) and all the long chains 
another concentration, $Z(t)=y_{m,n}(t)$.  
With these assumptions the very large system of rate equations 
(\ref{shortdot})--(\ref{longdot}) reduces to a coupled pair of 
ordinary differential equations 
\beqa
\dot Y & = & 2 \eta N Z + \epSplus{\lambda} \nn\\&& 
- 2 N Y \epLplus{\Lambda} \lbl{x1uniYdot} \\
\dot Z & = & - \eta Z + 2 Y \epLplus{\Lambda} 
\lbl{x1uniZdot} \eeqa
It is not possible to find a steady-state solution 
of these equations, since setting $\dot Z=0$ in 
equation (\ref{x1uniZdot}) implies that the flux from
long to short chains caused by hydrolysis exactly 
balances the growth of long from short.  Such a 
balance in (\ref{x1uniYdot}) in turn implies that $\dot Y$ 
must be strictly positive, removing any possibility 
of a physically relevant equilibrium configuration 
existing.   Indeed, as this system evolves, the 
concentration of long chains grows without bound, 
and the concentration of short chains decreases to 
zero.  The catalytic properties of chain growth 
dominate the system, leaving hydrolysis as a negligible 
effect, to the extent that there is a finite time 
singularity.  The leading order balance in the first 
equation (\ref{x1uniYdot}) is $2^\lambda \zetaSL^\lambda 
N^{3\lambda} Z^{2\lambda} \sim 2 Y \zetaLL^\Lambda 
N^{2\Lambda} Z^{2\Lambda}$, implying that as 
$Z\rightarrow\infty$, $Y\sim1/\surd Z$.  Then 
(\ref{x1uniZdot}) implies $\dot Z \sim Z^{2\lambda}$, 
and so $Z \sim (1-t/t_c)^{1/(2\lambda-1)}$.  The strong 
catalytic properties of long chains coupled with 
irreversible polymerization reactions cause the rapid 
growth in the concentration of long chains to occur.   
If we were to introduce reversible reactions 
($\beta_2\neq0\neq\beta_3$) then the singularity would 
be removed.  This modification would also cause a 
uniform steady-state to appear, in which all the 
concentrations of short chains were equal, and all the 
concentrations of long chains were equal. 

A singularity of the form described above is 
also present in the full system of equations 
(\ref{shortdot})--(\ref{longdot}) with $\beta_2=0=\beta_3$. 
This case is thus rather unphysical, so we shall 
turn to the situation where density is conserved 
-- an assumption which is more physically 
relevant and will also eliminate the singularity 
encountered above. 

\subsection{Constant total concentration of ribonucleotides in the soup}
\label{sec62}

The full kinetic equations for the case where all the 
catalytic growth reactions are irreversible ($\beta_2=\beta_3=0$) 
and density ($\ro$) is kept constant are 
\beqa \hspace*{-6mm}
\dot y_{m,n} & \!\!=\!\! & \!- \eta y_{m,n} + ( y_m \!+\! y_n ) 
\left( \ro \!-\! \Lambda \psum y_p \!-\! 2 \Lambda \pqsum y_{p,q} \right)^\Lambda 
\!\left( \epL \!+\! \alphaL y_{m,n} \!+\! \pqsum \chiL y_{p,q} \!+\! 
\!\pqsum \zetaLL y_{p,q} y_{m,n} \!\right)^\Lambda , 
\nn\\&&\lbl{62full1}\\ 
\hspace*{-6mm} \dot y_n\! & \!\!=\!\! & 
\msum \eta ( y_{m,n}\!+\!y_{n,m} ) \lbl{62full2}
\\&& \hspace*{-6mm}- \msum y_n 
\left( \ro - \Lambda \psum y_p - 2 \Lambda \pqsum y_{p,q} \right)^\Lambda 
\left( \epL + \alphaL y_{m,n} + \pqsum \chiL y_{p,q} + 
\pqsum \zetaLL y_{p,q} y_{m,n} \right)^\Lambda 
\nn\\&& \hspace*{-6mm}- \msum y_n 
\left( \ro - \Lambda \psum y_p - 2 \Lambda \pqsum y_{p,q} \right)^\Lambda 
\left( \epL + \alphaL y_{n,m} + \pqsum \chiL y_{p,q} + 
\pqsum \zetaLL y_{p,q} y_{n,m} \right)^\Lambda \nn\\&& \hspace*{-15mm} + 
\left( \!\ro \!-\! \Lambda \psum y_p \!-\! 2 \Lambda \pqsum y_{p,q}
\!\right)^{\!\Lambda}\!\!\left( \!\epS \!\!+\! \alphaS y_n \!+\! \msum \!\chiS
y_m \!+\!\! \pqsum \!\chiX y_{p,q} \!+\!\! \pqsum \!\zetaSS y_{p,q} y_n\!+\!\! 
\mpqsum \!\zetaSL y_{p,q} (y_{m,n}\!\!+\!y_{n,m}) \right)^{\!\lambda\!} ,
\hspace*{-4mm}\nn\eeqa
where it will be recalled that $N=4^{\Lambda}$, the number of
sequences of length $\Lambda$.   Of course, even this set of 
coupled equations cannot be solved exactly either analytically 
(due to its nonlinearity), or numerically (since it is NP hard). 
However, we can make important progress by using theoretical 
methods to study certain special solutions and their local behaviour. 

\subsubsection{Approach to the equilibrium solution}

The simplest solutions of all are the time-independent ones, 
and in this case there is a family of equilibrium solutions. These 
equilibrium solutions are degenerate, in that only short 
chains are present; there are no monomers or long chains.  
It is thus a genuine equilibrium state since there are no fluxes 
present in the system.   The reason for the time-independent solution 
being a true equilibrium solution rather than a non-equilibrium
steady-state is that once 
all monomers have been converted into chains (whether short or 
long), no further growth can occur; the only remaining process is 
hydrolysis, which converts all the long chains into shorter oligomers. 
Hence as $t\rightarrow\infty$, the system approaches 
\beq
\ol{y}_{m,n} = 0 , \hspace{8mm} 
\ol{x}_1=0 , \hspace{8mm} 
\ol{y}_m \geq 0 \;\;\;\; \mbox{ such that } \;\;\;\;
\msum \ol{y}_m = \ro / \Lambda , 
\eeq
which is an arbitrary distribution of short chains satisfying 
mass conservation.  Thus the kinetics allowed for in the model 
permit highly non-uniform equilibrium solutions, including the 
extinction of species (that is all the longer chains, and any 
number of the shorter chains). 

Expanding the dynamical behaviour about the equilibrium solution, 
we can find the late-time asymptotics which tell us how such a 
self-reproducing soup of ribonucleotides approaches an equilibrium 
state.   The quantities $y_{m,n}(t),x_1(t)$ both tend to zero, 
and we put $y_n(t) = \ol{y}_n - h_n(t)$ with $h_n(t)\rightarrow0$ 
as $t\rightarrow\infty$.  To leading order, 
the quantities $h_n,x_1,y_{m,n}$ then satisfy 
\beqa
x_1 & = & \Lambda \left( \qsum h_q - 2 \pqsum y_{p,q} \right) ,\lbl{apeqx1}\\ 
\dot y_{m,n} & = & - \eta y_{m,n} + \epL^\Lambda \Lambda^\Lambda 
(\ol{y}_m\!+\!\ol{y}_n) \left( \qsum h_q - \pqsum y_{p,q} \right)^\Lambda ,
\lbl{apeqymn} \\
\dot h_n & = & - \eta \msum ( y_{m,n} \!+\! y_{n,m} ) - \Lambda^\Lambda
\left( \qsum h_q - 2 \pqsum y_{p,q} \right)^\Lambda \left[
\left( \epS \!+\! \alphaS \ol{y}_n \!+\! \frac{\chiS\ro}{\Lambda} \right)^\lambda 
- 2 N \epL^\Lambda \ol{y}_n \right] . \nn\\&& \lbl{apeqhn}
\eeqa

As in our earlier study of self-reproducing vesicles~\cite{cw98}, 
the phase space trajectories followed by the mathematical 
solutions of these kinetic equations are not strongly attracted 
to an equilibrium configuration; rather the dynamics close to 
equilibrium occurs on a {\em centre manifold},  where nonlinear 
terms dominate the behaviour.    Motion on this manifold occurs 
slowly: relaxation does not follow an exponential decay with a 
characteristic timescale (relaxation time); rather the concentrations 
vary as inverse powers of time, with exponents $-1/\lambda$ 
for short chains and monomers,  and $-\Lambda/\lambda$ for 
long chains.   

To explicitly find the nature of the approach to equilibrium, 
we substitute 
\beq
x_1(t) = \tilde{x}_1 t^{-1/\lambda} , \hspace{8mm} 
h_n(t) = \tilde{h}_n t^{-1/\lambda} , \hspace{8mm}  
y_{m,n}(t)=\tilde{y}_{m,n} t^{-\Lambda/\lambda} , 
\lbl{62CM} \eeq
into equations (\ref{apeqx1})--(\ref{apeqhn}).  This yields 
\beq
\tilde{x}_1 = \Lambda \tilde{H} , \hspace{9mm} 
\tilde{y}_{m,n} = \frac{ \epL^\Lambda \Lambda^\Lambda 
\tilde{H}^\Lambda (\ol{y}_m\!+\!\ol{y}_n)}{\eta} , \hspace{9mm} 
\tilde{h}_n = \lambda \Lambda^\Lambda \tilde{H}^\Lambda 
\left[ \frac{2\ro\epL^\Lambda}{\Lambda} + \left( \epS \!+\! 
\alphaS \ol{y}_n \!+\! \frac{\ro\chiS}{\Lambda} 
\right)^\lambda \right] . \hspace*{-7mm} \lbl{62CMcoeff} \hspace*{-7mm}
\eeq
where $\tilde{H} = \qsum \tilde{h}_q$. The latter expression 
enables $\tilde{H}$ to be found:
\beq
\frac{1}{\tilde{H}^\lambda} = \lambda \Lambda^\Lambda \left[ 
\frac{ 2 \epL^\Lambda \ro N}{\Lambda} + \qsum \left( 
\epS + \alphaS \ol{y}_q + \frac{\ro\chiS}{\Lambda} 
\right)^\lambda \right] . \lbl{62CMH}
\eeq

These results show that the short chains with largest 
equilibrium concentration $\ol{y}_n$ also have the 
fastest growth rates near equilibrium, since the larger 
$\ol{y}_n$, the larger $\tilde{h}_n$.   Note, however, 
that this is only a {\em local} analysis: we have only 
found how the system behaves once it is close to any 
one of its degenerate equilibrium solutions; we have 
not yet discovered anything about which equilibrium solution 
is approached, or how it evolves from an initial 
condition towards equilibrium.   To gain more insight 
about earlier times in the evolution of the system, 
we now analyse a special solution--the uniform solution. 

\subsubsection{Uniform solution and its stability}
\label{stab-uni}

We now turn to analyse more general far from 
equilibrium kinetics; in particular, we shall be concerned
with investigating whether the uniform growth 
of all chains is a stable configuration. 
The {\em Ansatz} 
\beq
y_{m,n} = Z(t) , \hspace{9mm} y_m(t)=Y(t) , \lbl{Ansatz}
\eeq
when inserted into (\ref{62full1})--(\ref{62full2}), implies 
\beqa
\dot Z & = & - \eta Z + 
2 Y \rominus{\Lambda} \epLplus{\Lambda} \lbl{uni1} \\
\dot Y & = & 2 \eta N Z - 2 N Y \rominus{\Lambda} \epLplus{\Lambda}
+\nn\\&&+ \rominus{\Lambda} \epSplus{\lambda} . \lbl{uni2}
\eeqa

These equations are not explicitly solvable in terms of 
elementary functions due to the number and strength 
of the nonlinear terms present.  The {\em Ansatz} has, 
however, removed the NP hard complexity of equations 
(\ref{62full1})--(\ref{62full2}) so, if desired, a numerical 
solution of (\ref{uni1})--(\ref{uni2}) could be found 
using standard methods.  We shall not pursue this here, 
as it is our aim to show that the solution is in fact unstable 
and so will not be realised. The presence of an instability 
means that any small perturbation of the solution will grow 
as time progresses, and the full system (\ref{62full1})--(\ref{62full2}) 
will not approach the equilibrium solution of 
(\ref{uni1})--(\ref{uni2}).  In our system of equations 
we shall assume that the instability is initiated by 
slightly nonuniform initial conditions; in a more general 
model the instability could alternatively be caused by 
different chain types having slightly different reaction 
rates.  We shall also demonstrate that the instability is 
effective at earlier times for certain choices of parameters, 
indicating that the uniform solution will not be 
manifested in the solution of (\ref{62full1})--(\ref{62full2}). 

To illustrate this instability mathematically, we perform 
a linear expansion about the uniform solution, putting 
$y_{m,n} = Z + Z \hat y_{m,n}$, $y_n = Y + Y \hat y_n$, 
with $\hat y_n, \hat y_{m,n} \ll 1$, into 
(\ref{62full1})--(\ref{62full2}).  At leading order, this produces 
\beqa
\dot{\hat{y}}_n & = & A(t) \hat{y}_n + B(t) 
\msum (\hat{y}_{m,n}\!+\!\hat{y}_{n,m} ) \lbl{624stab1} \\
\dot{\hat{y}}_{m,n} & = & C(t) \hat{y}_{m,n} + 
D(t) [ \hat{y}_m + \hat{y}_n ] . \lbl{624stab2}
\eeqa
where 
\beqa
A(t) & \!=\! & - \frac{2N\eta Z}{Y} + 
\frac{1}{Y} \left[ (\lambda\!-\!1) (\alphaS+N^2\zetaSS Z) Y 
- ( \epS + N\chiS Y + N^2\chiX Z + 2 \zetaSL N^3 Z^2 ) \right] 
\times\nn\\&&\times \rominus{\Lambda} \epSplus{\lambda-1} \nn\\
B(t) & \!=\! & \frac{\eta Z}{Y} - 
\Lambda Z (\alphaL\!+\! \zetaLL N^2 Z) \rominus{\Lambda} 
\epLplus{\lambda} +  \nn \\&& +
\frac{\lambda \zetaSL N^2 Z^2}{Y}
\rominus{\Lambda} \epSplus{\lambda-1}  \nn\\
C(t) & \!=\! & \frac{2Y}{Z} \rominus{\Lambda} \epLplus{\lambda}
\left[ \lambda Z ( \alphaL\!+\!\zetaLL N^2 Z ) 
- ( \epL\!+\!\chiL N^2 Z) \right] \nn\\
D(t) & \!=\! & \frac{Y}{Z} \rominus{\Lambda} \epLplus{\Lambda} . 
\lbl{ABCDdef} \eeqa
Now it can be shown that if $\msum\hat{y}_m=0=\pqsum \hat y_{p,q}$ 
at any instant in time, then these quantities are zero for all time. 
Since it is sufficient to consider perturbations of such a form, 
this result has been used to simplify the above expressions.   

Since equations (\ref{624stab1})--(\ref{624stab2}) are linear 
in $\hat{y}_n,\hat{y}_{m,n}$, the temporal dimension of the 
problem can be separated from the compositional part 
by expressing the solution in the form 
\beq
\hat{y}_{n}(t) = \xi_n S(t) , \hspace{9mm} 
\hat{y}_{m,n}(t) = \xi_{m,n} L(t) .  \lbl{separation}
\eeq
This approach introduces a new parameter into the problem -- the 
separation constant $K$, where $\xi_{m,n} = K (\xi_m + \xi_n)$.  
Proving that the uniform solution is unstable then reduces to 
demonstrating that the functions $L(t),S(t)$ grow as time ($t$)
increases.  These functions satisfy the differential equations 
\beq
\left( \begin{array}{c} \dot S \\ \dot L \end{array} \right) = 
\left( \begin{array}{cc} A(t) & 2NKB(t) \\ D(t)/K & C(t) \end{array} \right) 
\left( \begin{array}{c} S \\ L \end{array} \right) . 
\lbl{stabmat} \eeq

{}From this it is possible to deduce that the uniform solution is 
unstable for certain choices of the parameters.  A detailed analysis 
is hindered by the fact that the uniform solution for $Y(t),Z(t)$ 
(\ref{uni1})--(\ref{uni2}) cannot be explicitly determined. 

The simplest part of the reaction kinetics to analyse is the 
approach to equilibrium, and this is where the most rigorous 
results can be obtained.  We shall discuss this stage of the 
reaction first, and then proceed to derive more general and 
approximate results for earlier stages of the polymerisation scheme.  

\subsubsection{Instability in the approach to equilibrium}
\lbl{approach-eqm}

The only equilibrium solution of the system 
(\ref{uni1})--(\ref{uni2}) is  $\ol{Z}=0$, 
$\ol{Y}=\ro/N\Lambda$ which implies $\ol{x}_1=0$. 
The approach to this state is governed by 
\beq
x_1(t) \sim \Lambda N Y_1 t^{-1/\lambda} , \hspace{7mm} 
Y(t) \sim \frac{\ro}{\Lambda N} - Y_1 t^{-1/\lambda} , \hspace{7mm} 
Z(t) \sim Z_1 t^{-\Lambda/\lambda} , \hspace{4mm} 
\mbox{ as $t\rightarrow\infty$} ,  \lbl{623asy}
\eeq
where $Y_1,Z_1>0$ can be found either from an expansion of 
(\ref{uni1})--(\ref{uni2}) about the equilibrium solution, or by 
inserting the {\em Ansatz} (\ref{Ansatz}) into 
(\ref{62CMcoeff})--(\ref{62CMH});  they are thus determined by 
\beq
\frac{1}{Y_1^\lambda} = 
\lambda \Lambda^\lambda N^\Lambda \left\{ 
2\ro\epL^\Lambda + \Lambda \left( 
\epS + \frac{\ro\alphaS}{\Lambda N} +
\frac{\ro\chiS}{\Lambda} \right)^\Lambda \right\} , 
\hspace{9mm} 
Z_1 = \frac{ 2 \ro \Lambda^\lambda N^\lambda 
\epL^\Lambda Y_1^\Lambda }{\eta} . 
\eeq

An examination of (\ref{624stab1})--(\ref{ABCDdef}) 
in the limit $t\rightarrow\infty$ highlights one 
condition for an instability to exist. 
In this limit, the kinetic equations simplify to 
\beq
t^{\Lambda/\lambda} \dot{\hat{y}}_n = 
A_0 \hat {y}_n + 2 ( \Lambda N Y_1 \epL )^\Lambda 
\msum ( \hat{y}_{m,n}\!+\!\hat{y}_{n,m} ) ,  
\hspace{9mm} 
\dot{\hat{y}}_{m,n} = \half \eta \left( 
\hat{y}_m + \hat{y}_n - \hat{y}_{m,n} \right) ,  
\eeq
where 
\beq
A_0 = (N\Lambda Y_1)^\Lambda \left\{ \left[ 
(\lambda\!-\!1)\alphaS - N \left( 
\frac{\Lambda\epS}{\ro} + \chiS \right) \right] 
\left( \epS + \frac{\alphaS \ro}{N\Lambda} + 
\frac{\chiS\ro}{\Lambda} \right)^{\lambda-1} - 
4 N \epL^\Lambda \right\} . 
\eeq
Separating the temporal and compositional variables, 
as in (\ref{separation}), we obtain a simplified 
system of equations which can be written as 
\beq
t^{\Lambda/\lambda} \frac{d^2 L}{dt^2} + 
( \eta t^{\Lambda/\lambda} - A_0 ) \frac{dL}{dt} -
\eta [ A_0 + 2 N (\epL\Lambda NY_1)^\Lambda ] L = 0 . 
\lbl{prekummer} \eeq

Unfortunately it is not possible to solve this 
equation analytically, but the closely related system 
\beq 
t \frac{d^2 L}{dt^2} + ( \eta t - A_0 ) \frac{dL}{dt} -  
\eta ( A_0 + 2 N (\epL\Lambda NY_1)^\Lambda ) L = 0 , 
\lbl{kummereq}  \eeq
can be solved in terms of hypergeometric functions (see 
Abramowitz \& Stegun \cite{as72} for details). 
This approximation is easily justified by noting 
that the range of chain lengths we are interested 
in is $\Lambda=\lambda\!+\!1$ to be approximately 50.  
The solution of (\ref{kummereq}) can be written as 
\beq  
L(t) = K_1 M \left( -2N (\epL\Lambda NY_1)^\Lambda - A_0 , 
-A_0 , -\eta t \right) + K_2 U \left( 
-2N (\epL\Lambda NY_1)^\Lambda - A_0 , -A_0 , -\eta t \right) , 
\eeq
where $K_1,K_2$ are constants of integration and the 
functions $M(\cdot),U(\cdot)$ are as defined in Chapter 13 of 
Abramowitz \& Stegun \cite{as72}.  This solution grows in the limit 
$t\rightarrow\infty$ if $A_0 + 2 N (\epL\Lambda NY_1)^\Lambda>0$. 
This condition can be rewritten as  
\beq
\left[ (\lambda\!-\!1) \alphaS - N \left( 
\frac{\epS\Lambda}{\ro}+\chiS\right) \right] 
\left( \epS + \frac{\alphaS\ro}{N\Lambda} +
\frac{\chiS\ro}{\Lambda} \right)^\Lambda 
> 2 N \epL^\Lambda . \lbl{623ineq}
\eeq
Thus the instability depends on the autocatalytic rate 
being greater than the sum of the uncatalysed rate and 
crosscatalysed (copying-error) rate by a factor 
which depends on the number of chains.  

Qualitatively, this agrees with our intuition: in order 
for perturbations to grow, {\em the autocatalytic rate should 
be greater than the total rate at which error-prone copies 
are produced} (whether that be by the uncatalysed or the 
catalysed mechanism). 

\subsubsection{More general instability analysis of the uniform solution}

Formally, the Routh-Hourwitz criteria \cite{Murray} for 
(\ref{stabmat}) are that the system is stable if $A+C<0$, 
and $AC>2NBD$. Therefore, to prove that the system is 
unstable our aim is to prove at least one of 
\beq
A + C > 0 , \hspace{9mm} AC < 2NBD \lbl{RHineq}
\eeq
holds.  However, since $A,B,C,D$ are functions of time ($t$), 
results based on these inequalities rely on interpreting 
stability in the looser manner of `as $t$ increases the 
perturbation grows', rather than a strict argument in the 
style of `as $t\rightarrow\infty$, $S(t),L(t) \sim f(t)$, for 
some given $f(t)$'.  More rigorous results in the limit 
$t\rightarrow\infty$ have already been given in Section 
\ref{approach-eqm}. 

One interesting point from a mathematical perspective is that the 
stability criteria do not depend on the separation parameter $K$.  
Thus the bifurcation where the transition from stable to unstable 
behaviour occurs is degenerate in that  many modes of the system 
change stability at the same point in parameter space. 

Routh-Hourwitz analysis confirms the presence of an instability 
as the system (\ref{stabmat}) approaches equilibrium, but since 
we are now working with a slightly different definition of 
stability and have not made the approximation by which 
(\ref{prekummer}) is replaced by (\ref{kummereq}), we obtain a 
slightly different inequality 
\beq
\alphaS > \frac{ N (\Lambda\epS+\ro\chiS)}{(\lambda-1) \ro} . 
\lbl{624ineq} \eeq
This inequality can be interpreted in a number of different 
ways; for example, it sets upper limits on the uncatalysed 
growth rates for chains, $\epS$. Another interpretation 
is that it prescribes an upper limit on the error-prone 
self-replication rates $\chiS$. The limits, however depend 
on the chain length ($N$), so a third 
explanation is that it sets a maximum on $N$, the length of 
chain which can successfully self-replicate and maintain its 
concentration at a significantly higher level than other chains 
of the same length. Finally, it prescribes a minimum 
for the autocatalytic rate constant above which the differences 
in concentration between chains grow. 

We progress now to apply the inequalities (\ref{RHineq}) to 
special cases of the system (\ref{624stab1})--(\ref{ABCDdef}), 
since it is then simpler to draw more specific conclusions.  
The special cases we consider are (i) systems with only 
crosscatalysis and autocatalysis, and (ii) systems where 
ribozymic synthesis dominates chain polymerisation. 
Of course in the RNA world both processes occur 
simultaneously and interact with each other; only an 
analysis of the full system (\ref{624stab1})--(\ref{ABCDdef}) 
will reveal how these mechanisms affect each other, but some 
useful preliminary insight and results are gained from 
analysing the processes independently. 

\subsubsection{Effect of template-based catalysis on the uniform solution}
\lbl{625-sec}

To analyse the effect of standard template-based synthesis 
on the process, we set all the other rate constants to zero,  
$\epL=\epS=\zetaSS=\zetaSL=\zetaLL=0$ leaving only 
$\alphaL,\alphaS,\chiS,\chiX,\chiL,\eta\neq0$. We then 
solve the model subject to initial conditions in which only 
a few chains are present ($0<Y(0),Z(0)\ll1$).  
The condition $A+C>0$ can then be simplified to 
\beqa \hspace*{-5mm}
\left( \alphaS Y\!+\!N\chiS Y\!+\!N^2\chiX Z\right)^{\lambda-1} 
\left[ (\lambda\!-\!1)\alphaS Y\!-\!N\chiS Y\!-\!N^2\chiX Z\right]
+&&\lbl{625ineq1}\\ +2 Y^2 Z^\lambda (\alphaL+N^2\chiL)^\lambda 
[\lambda\alphaL-N^2\chiL] &>& \frac{2N\eta Z}{\rominus{\Lambda}} . 
\nn \eeqa
This inequality is highly suggestive of the simple conditions 
\beq
(\lambda-1) \alphaS > N \chiS , \hspace{8mm} 
\lambda \alphaL > N \chiL . \lbl{simplecond}
\eeq
Since the right-hand side of (\ref{625ineq1}) is positive, 
one or both of the terms on the left-hand side must be too. 
Requiring the first term to be positive implies the 
first part of (\ref{simplecond}) and the second term 
in (\ref{625ineq1}) yields the second. 
If both parts of equation (\ref{simplecond}) hold then 
for small enough $\eta$ an instability will exist.   
Such inequalities are fully 
in accord with earlier results ((\ref{623ineq}) and 
(\ref{624ineq})) as well as our intuition that 
autocatalysis should dominate crosscatalysis in order for 
distinct species of chains to emerge and remain viable. 

The other condition ($AC<2NBD$) can be rewritten as 
\beqa  \lefteqn{ 
\left[ \lambda \alphaL \!-\! N^2 \chiL \right] \left[ 
(\lambda\!-\!1) \alphaS Y \!-\! N\chiS Y \!-\! N^2\chiX Z \right] 
\left( \alphaS Y\!+\!N\chiS Y \!+\! N^2 \chiX Z \right)^{\lambda-1}
+\hspace*{50mm}}&&\hspace*{12cm}\lbl{625ineq2}\\&&\hspace*{50mm}+\; 
N\Lambda\alphaL YZ^\Lambda(\alphaL\!+\!N^2\chiL)^\Lambda \;\; < \;\;  
\frac{N\eta Z\left[(\Lambda\!+\!\lambda)\alphaL-N^2\chiL\right]}
{\rominus{\Lambda}} . \nn 
\eeqa
which shows that the presence of hydrolysis 
(the $\eta$ term) can cause an instability.   
Note that while the previous inequality gave 
an upper limit for $\eta$, this inequality 
prescribes a lower bound.  At this stage, 
it is perhaps worth repeating that only one 
of the two inequalities (\ref{625ineq1}) and 
(\ref{625ineq2}) need be satisfied for an 
instability in the system to exist. 

Simple inequalities such as (\ref{624ineq}) 
and (\ref{simplecond}) are actually quite 
stringent; since $N$ is exponentially large, 
the autocatalytic reaction rate must far exceed 
that of the crosscatalytic rate.  It is helpful 
to define a quality factor $Q$ for these 
template-based chain synthesis mechanisms. 
There are $N$ crosscatalytic processes in our 
model, each operating at the rate $\chi$, only 
one of which produces an accurate copy of the 
catalyst,  and there is one autocatalytic 
mechanism which is taken to provide an exact 
replication of the catalyst, at rate $\alpha$.  
Thus from all the template-based catalytic chain 
synthesis processes the proportion of accurate copies is 
\beq
Q = \frac{\alpha+\chi}{\alpha+N\chi} . 
\eeq
This quantity can be defined for the formation 
of short or long chains.  The inequalities 
(\ref{624ineq}) and (\ref{simplecond}) can be 
recast into conditions on $Q$, in the manner of 
Nu\~{n}o \etal\ \cite{ne1}, and in this form, they 
do not appear so demanding. For example, the 
second of the inequalities in (\ref{simplecond}) becomes
\beq
 Q > \frac{1}{\Lambda} + \frac{\lambda}{\Lambda N} . 
\eeq

By considering these template-based catalytic 
processes along with uncatalysed RNA polymerization 
(that is, ignoring ribozymic synthesis) it is possible 
to make a crude estimate of the timescales over which 
RNA chain formation occurs.  Integrating a simplified 
but dimensional version of (\ref{x1uniYdot})
\beq
\dot Y = \ro^\Lambda (\epS+\alphaS Y)^\lambda , 
\eeq
we find 
\beq
t \approx \frac{1}{ (\lambda\!-\!1)  \ro^\Lambda 
\epS^{\lambda-1} \alphaS } . \lbl{tlong}
\eeq
Taking a chain length of $\Lambda=10$, a concentration 
(density) of $\ro=10^{-3}$M and $\alpha=10^5\epS$, we 
obtain for the possible number of chain sequences $N=10^6$.   
The timescale (\ref{tlong}) then gives a quantity which 
is consistent with RNA chains forming in months--years, 
if the value chosen for $\epS$ is taken to lie in the  
range $3.5-4.5$ M$^{-1}$sec$^{-1/\lambda}$. 
A much faster timescale is obtained if one calculates 
the rate of divergence from the uniform solution 
according to $A(t)$ in equation (\ref{624stab1}).  
{\em Thus the model shows that the selection of one 
species over another happens on a much faster timescale 
than the growth of chains.}  
These estimates are crude due to their reliance on 
raising the reaction rate $\epS$ and density $\ro$ 
to the power $\Lambda$; for example, with $\Lambda=10$, 
an alteration of a parameter by a factor of two yields 
a factor of $2^{10}\approx 10^3$ in the final 
approximation for the timescale. 

\subsubsection{Effect of ribozymic synthesis on the uniform solution}

It is harder to derive concise results from the case of 
ribozymic synthesis since these mechanisms introduce 
higher nonlinearities and coupling between different 
chains than the usual catalytic mechanisms. 

In this case we set all catalytic and normal rate constants 
to zero ($\epL=\epS=\alphaL=\alphaS=\chiS=\chiX=\chiL=0$) and 
keep only the hydrolysis and ribozymic synthesis rates nonzero 
($\eta,\zetaSS,\zetaSL,\zetaLL$), cf.~Table \ref{6tab-inc}.  The 
condition $A+C>0$ (equation (\ref{RHineq})) then reduces to 
\beqa
2\lambda \zetaLL^\Lambda N^2 Y^2 Z^{\Lambda+1} \;+\; \left( 
\zetaSS Y \!+\! 2 \zetaSL N Z \right)^{\lambda-1} \,\left[\, 
(\lambda\!-\!1) \zetaSS Y - 2 \zetaSL N Z \,\right] & > &
\frac{2\,\eta\,N^{1-2\lambda}\,Z^{1-\lambda}}{\rominus{\Lambda}} , 
\nn\\&& \lbl{626ineq1} \eeqa
which clearly fails when $Z$ is small and when the monomer 
concentration is small, since in both these cases the 
right-hand side becomes large.  However at intermediate 
times this inequality can be satisfied.  For example, if 
$\zetaSS$ is sufficiently larger than $\zetaSL$, then the 
second term could become large enough to dominate the 
right-hand side. This condition shows that the highly 
specific nature of ribozyme-assisted synthesis of a 
short chain by an identical or complimentary short chain 
is of greater importance to the instability than the 
less selective synthesis of a short chain by 
a long chain.   Alternatively, the first term indicates 
that a large value of $\zetaLL$ (ribozyme-assisted synthesis 
of long chains by long chains) can instigate the instability. 

The other instability condition $AC<2NBD$ reduces to 
\beqa  \lefteqn{
\frac{\lambda}{N}\left(\zetaSS Y+2\zetaSL NZ\right)^{\lambda-1} 
\left[ (\lambda\!-\!1)\zetaSS Y - 2 \zetaSL N Z \right] + } \\ 
\hspace*{2cm} && + \; N^2\Lambda \zetaLL^\Lambda Y Z^{\Lambda+1} 
\;\; < \;\; 
\frac{\eta\,(\Lambda+\lambda)\,N^{-2\lambda}\,Z^{1-\lambda}}
{\rominus{\Lambda}} \;+\; 
\lambda\zetaSL Z\left(\zetaSS Y\!+\!2\zetaSL NZ\right)^{\lambda-1}.
\nn \eeqa
In the limits $Z\rightarrow0$ and $x_1\rightarrow0$ this 
inequality is satisfied, implying the system is 
unstable in these limits even though the previous 
inequality (\ref{626ineq1}) is not satisfied. 
Due to the large size of $N$, we expect $N^{-2\lambda}$ 
to be insignificant, and so if this inequality were to 
hold, it would be mainly due to the presence of 
ribozyme-assisted synthesis of short chains by long 
-- the $\zetaSL$ term. 


\section{Discussion} \label{discuss}
\setcounter{equation}{0}

The question of which type of RNA chain (one ribonucleotide sequence 
or a set of such sequences) will grow from an initial soup of 
monomers presents an extremely difficult challenge both 
experimentally and theoretically.    From the latter perspective 
with which we have been concerned in this paper, 
it is a massively degenerate version of the problem of polymorphism, 
that is of determining which particular crystal structure among several 
will grow from a melt or supersaturated solution.  This problem 
is known to depend on the kinetics of the process--it is not an 
equilibrium phenomena: the thermodynamically most stable macroscopic 
structure is not necessarily the one which nucleates first or grows 
the fastest, as is well known in the fields of biomineralisation 
\cite{falini} and the crystallisation of macromolecules 
\cite{kam,beckmann}. In the present context, the difficulty is 
compounded many times over owing to the combinatorial complexity 
of polyribonucleotide self-reproduction, which causes the problem 
to acquire an NP hard character in the language of algorithmic 
complexity theory~\cite{ch95}.  

The models we have introduced and analysed in this paper 
include many catalytic mechanisms  
which affect the rate at which various reactions proceed; these include
{\em inter alia} template-based and ribozymal replicase-assisted 
RNA synthesis. Our paper has addressed a basic question concerning
the putative origins of the RNA world from the
standpoint of plausible chemical kinetics and provides important
evidence supporting the notion that self-reproducing ribonucleotide
polymers could quite easily have emerged from a prebiotic soup. In 
reaching this conclusion, we have analysed carefully the 
nonlinearities with which this problem is replete. It is worth
pointing out that much of the previous work on related issues,
including that of Eigen \etal\ on 
hypercycles~\cite{eigen,eigenetal,eigmccsch,ems88} was based on the
analysis of {\em linearised} kinetic equations, a procedure that 
drastically suppresses the difficulty of the problem together 
with the richness of possible behaviour. Nu\~{n}o \etal 's work, 
like that of Eigen \etal, also assumes the existence of 
self-reproducing hypercycles, but does incorporate certain 
nonlinearities~\cite{ne1,ne2}. 
Addressing a different problem, they model template-based 
synthesis using quadratic nonlinearities as in our models. 
In addition, our models, include ribozymically assisted 
synthesis described by cubic nonlinearities.  However, in our
approach the coarse-graining contraction procedure, used to 
overcome the combinatorial explosion of possible proliferating
polyribonucleotide sequences, introduces
extremely high nonlinearities into the analysis.  Overall, 
it is the key combination of {\em hydrolysis and catalysis} which we find
produces stable populations of a small number of selected chain types.  

Cross-catalysis (error-prone template-based copying) 
speeds up self-reproduction generally, 
but it tends to even 
out the distribution of sequences since it acts non-specifically.  
Autocatalysis and hydrolysis are 
sequence specific. Hydrolysis splits one chain into two: whereas a single 
long chain can catalyse only one reaction, the two shorter chains 
produced by hydrolysis can each catalyse reactions which make 
chains similar to the original long chain.  Thus hydrolysis aids the 
autocatalytic replication process by recycling and hence multiplying 
the number of chains of a given length capable of autocatalysis.  
This recycling of material and information can be thought of as 
primordial metabolism.  If hydrolysis acts more
slowly than autocatalysis there is not enough catalytic material 
around to maintain the instability driving the preferential formation
of certain sequences and the suppression of others; then 
other factors will even out the differences in concentration of
different species and no selectivity ensues. 

The results of Section \ref{red-hy-sec} show that even with 
hydrolysis, the presence of catalytic chain growth mechanisms 
implies that viable concentrations of long chains will eventually 
be formed if one is prepared to wait long enough. At the end 
of Section \ref{625-sec}, a simple calculation was carried out to 
estimate the length of this induction time.  Although estimates 
of numerical values of our parameters are extremely hard to find, 
the data available is consistent with a timescale of months for 
the formation of viable concentrations of chains of ten bases. 

In conclusion, we have demonstrated that it is possible to realise 
the selection of certain self-replicating RNA polymer chains 
in a reasonable amount of time starting from plausible 
assumptions  about the chemistry and initial conditions 
that could have prevailed within a putative prebiotic 
soup comprised of $\beta$-D-ribonucleotide monomers. 

\subsection*{Acknowledgments}

PVC is grateful to Pier Luigi Luisi for an invitation to visit his
laboratory at E.T.H., Z\"{u}rich, in June 1996 and for the two 
brainstorming sessions held there involving members of his 
group as well as Albert Eschenmoser. These lively and 
stimulating discussions were responsible for initiating the 
work in this paper. A further visit during May 1997 
helped to eliminate several misconceptions and thus to 
steer the paper into its present form.  
JADW is grateful to Wolfson College and the Department of 
Theoretical Physics, University of Oxford, for hosting 
discussions with PVC;  and acknowledges several helpful 
conversations with John King.  JADW wishes to thank both 
the Nuffield Foundation and the University of Nottingham 
for support under their New Lecturer's schemes. 


\newpage
\appendix
\renewcommand{\theequation}{\Alph{section}\arabic{equation}}
\section{Analysis of an alternative kinetic model for 
pure autocatalytic RNA replication without hydrolysis}
\lbl{appa}
\setcounter{equation}{0}

In Section~\ref{cross-sec}, we considered a detailed 
kinetic model for RNA replication that omitted hydrolysis 
and assumed that rate coefficients for polymerization 
were independent of chain length.   In this section we 
retain the same structure for the kinetic equations 
(\ref{ackin1}), but instead of assuming that the reaction 
rate coefficients are independent of chain length as 
in (\ref{ccJind}), here we make them proportional to 
chain length: $\alpha_{r,s}=rs$, $\alpha_{r,0}=\ep r$. 
The flux equations (\ref{ccJind}) are thus replaced by 
\beq
J_r^{\gamma,N_i}=c_r^\gamma c_1^{N_i}\left(
\ep r + \sum_\theta r s c_s^\theta\right), \hspace{5mm}
J_r^{N_i,\gamma}=c_r^\gamma c_1^{N_i}\left( 
\ep r + \sum_\theta r s c_s^\theta\right), \lbl{ccJdep} 
\eeq 
and inserted into the kinetic equations (\ref{ackin1}) 
(hydrolysis is again neglected).   We continue ignore 
the exact order of nucleotides in the chains ($\gamma$), 
concentrating purely on the length of chains, in order 
to investigate kinetics by which a distribution of long 
chains may be formed.  We define 
$f_r=\sum_{\gamma:|\gamma|=r}c_r^\gamma$ in the same 
sense as before, to simplify the governing equations
(\ref{ackin1}) and (\ref{ccJdep}) to 
\beq
\dot{f}_r = 
2 \fo f_{r-1} \left( \ep(r-1) + \sum_s (r-1) s f_s \right) 
- 2 \fo f_r \left( \ep r + \sum_s r s f_s \right) . 
\eeq
As in the previous example, the system of ordinary differential equations 
can be solved using the generating function (\ref{Fdef}), which now satisfies 
the partial differential equation 
\beq
\pad{F}{t} - 2\fo\pad{F}{z}(1-e^{-z})(\ep-G_0(t)) =
2\fo^2e^{-z}(\ep-G_0(t)) , 
\lbl{pde1} \eeq
where $G_0(t) = \pad{F}{z}|_{z=0}$ (thus $G_0(0)=-\fo$). 

Equation (\ref{pde1}) is significantly harder to solve than the previous 
example; it is best analysed using the method of characteristics.  First, 
however, we shall find the function $G_0(t)$.  We define 
$G(z,t)=\pad{F}{z}$ so that $G_0(t) = G(0,t)$ then (\ref{pde1}) implies 
$G(z,t)$ satisfies the equation 
\beq
\pad{G}{t} = 2\fo (1-e^{-z})(\ep-G_0) \pad{G}{z} - e^{-z} (\ep-G_0) G = 
- \fo e^{-z} (\ep-G_0) . 
\eeq
In the limit $z\rightarrow0$ this reduces to the ODE 
\beq
\frac{dG_0}{dt} = 2\fo \,[\, \ep - G_0(t) \,]\,[\, G_0(t)-\fo\,]  
\hspace{6mm} \Rightarrow \hspace{6mm} 
G_0(t) = \fo \left( \frac{ (\fo+\ep)e^{2t\fo(\fo-\ep)} - 2\ep } 
{ (\fo+\ep)e^{2t\fo(\fo-\ep)} - 2\fo } \right) , 
\eeq
when integrated using the initial condition $G_0(0)=-\fo$. This function 
gives us the first indication of singular behaviour, for it blows up when 
\beq
t = t_c := \frac{1}{2\fo(\fo-\ep)} \log\left(\frac{2\fo}{\fo+\ep}\right) 
\sim \frac{\log(2)}{2\fo^2} . \lbl{tcdef}
\eeq

To analyse the situation in more detail, we return to the full equation  
(\ref{pde1}) and solve the characteristic equations 
\beq \begin{array}{rclcrcl}
\ds\frac{dt}{d\sigma} & = & 1 & \Rightarrow & t & = & \sigma + h_1(\tau) \\
\ds\frac{dz}{d\sigma}
&=\!&\!-2\fo(1\!-\!e^{-z})(\ep\!-\!G_0(t))\!&\Rightarrow&
e^z&=&1+\exp \left(h_2(\tau)-2\fo\left[\,\ep\sigma-\int_0^\sigma G_0(s)ds\,
\right]\right)\\
\ds\frac{dF}{d\sigma} & = & 2 \fo^2 e^{-z} (\ep\!-\!G_0(t)) &
\Rightarrow&F&=&
h_3(\tau) + 2 \fo^2 \ds \int_0^\sigma \ds \frac{[\ep-G_0(s)] \, ds}
{ 1 + \exp[ h_2(\tau) - 2\fo (\ep s - \int_0^s G_0(p) dp ) ]}
\end{array} \eeq
Now we apply the initial conditions $F(z,0) = \fo e^{-z}$ 
on the line $\sigma=0$
\beq\begin{array}{rclcccrcl}
t & = & 0          && \Rightarrow && h_1(\tau) & = & 0              \\
z & = & \tau       && \Rightarrow && h_2(\tau) & = & \log(e^\tau-1) \\
F & = & \fo e^{-z} && \Rightarrow && h_3(\tau) & = & \fo e^{-\tau}  . 
\end{array}\eeq
It is possible to eliminate both $\sigma$ and $\tau$ to obtain the solution

\beq
F(z,t)=\frac{\fo}{1+(e^z-1)\exp(2\fo[\ep t - \int_0^t G_0(v)dv)])} +
\int_0^t 
\frac{ 2 \fo^2 [\ep-G_0(t-u)] \, du }{ 1 + (e^z-1) \exp(2\fo[\ep u -
\int_0^u
G_0(t-p) dp])} ; 
\eeq
however it is not obvious from this how the individual concentrations 
$f_r(t)$ behave.  We are in a similar position to that found in 
Section \ref{cross-sec}, and so we again treat the chain length distribution 
function $f_r(t)/F(0,t)$ as a time-dependent 
probability distribution function, extracting the number of chains, 
expected length and standard deviations from the function $F(z,t)$. 

The number density of chains, 
\beq
{\bf N}(t) = F(0,t) = \fo \, \left[ \, 1 + \log \left( \frac{(\fo-\ep)} 
{2\fo - (\fo+\ep) e^{2t\fo(\fo-\ep)}} \right) \, \right] \, ,  
\eeq
blows up at $t=t_c$ -- the same place as $G_0(t)$ became singular
(\ref{tcdef}).  To perform a local analysis we put $t=t_c-\tau$; then 
${\bf N}\sim-\fo\log\tau$, as $\tau\rightarrow0$. 
The expected chain length is 
\beq
{\bf E}(t) = \frac{-1}{F(0,t)} \pad{F}{z}(0,t) = \frac
{\{(\fo+\ep)e^{2t\fo(\fo-\ep)}-2\ep\}}
{\{2\fo-(\fo+\ep)e^{2t\fo(\fo-\ep)}\}} 
\frac{1}{\left[\, 1 + \log\left( \frac{(\fo-\ep)} 
{2\fo-(\fo+\ep)e^{2t\fo(\fo-\ep)}} \right) \,\right]} , 
\eeq
which also blows up at $t=t_c$.  Finally, the variance 
\beq
{\bf V}(t) = \err - \er^2 = 
\frac{1}{F(0,t)} \padd{F}{z}(0,t) - \frac{G_0(t)^2}{F(0,t)^2} , 
\eeq
for general $\ep$ is a very complicated function, but in 
the limit $\ep\rightarrow0$, it simplifies to 
\beq
{\bf V}(t) = \frac{ \left( 2 e^{2t\fo^2} - 1 \right) \left[\, 1 - \log
\left( 2 - e^{2t\fo^2} \right) \,\right] - e^{4t\fo^2} } {\left( 2 - e^{2t\fo^2} 
\right)^2 \, \left[\, 1 - \log \left( 2 - e^{2t\fo^2} \right) \,\right]^2 } . 
\eeq

Plotting ${\bf E}$ and $({\bf E\pm\sqrt{V}})$ against time for $0<t<t_c$, we find 
that the expected chain length (${\bf E}$) remains small for a long 
period of time, before increasing rapidly as $t\rightarrow t_c$, where 
it diverges.  ${\bf E+\sqrt{V}}$ behaves similarly although it always lies 
above ${\bf E}$; however ${\bf E-\sqrt{V}}$ diverges to $-\infty$ as 
$t\rightarrow t_c$, implying that shorter chains are 
suppressed by the kinetics. 
These results suggest that uniform growth is only mildly 
unstable during periods of slow increase in concentrations of long 
RNA chains, but highly unstable when rapid change occurs. 


\newpage 

\renewcommand{\baselinestretch}{1.2} 


\listoffigures

\listoftables

\newpage

FIGURE 1 

\begin{figure}[hbt] \vspace{3.5in} \begin{picture}(300,30)(-60,-10)
\thinlines
\put(00,00){\vector(1,0){175}}
\put(00,00){\vector(0,1){200}}
\put(50,50){\circle*{4}}
\put(50,-2){\line(0,1){3}}
\put(-2,50){\line(1,0){3}}
\put(120,-2){\line(0,1){3}}
\put(-2,150){\line(1,0){3}}
\put(45,-12){$c_1$}
\put(-15,45){$c_1$}
\put(175,-12){$C_0$}
\put(-15,200){$u_0$}
\put(115,-16){$\ol{C}_0$}
\put(-15,145){$\ol{u}_0$}
\put(55,40){IC}
\thicklines
\put(50,50){\vector(2,1){10}}
\put(52,51){\line(2,1){10}}
\put(120,150){\circle*{4}}
\put( 70,125){\vector(2,1){20}}
\put( 90,135){\vector(2,1){10}}
\put(100,140){\line(2,1){40}}
\put(170,175){\vector(-2,-1){20}}
\put(150,165){\vector(-2,-1){10}}
\put(175,175){$W_S$}
\put(105,120){\vector(1,2){10}}
\put(110,130){\vector(1,2){20}}
\put(135,180){\line(-1,-2){10}}
\put(110,110){$W_C$}
\put(71,62){\circle*{1}}
\put(82,75){\circle*{1}}
\put(90,90){\circle*{1}}
\put(98,105){\circle*{1}}
\end{picture} \end{figure}

\newpage

FIGURE 2a

\begin{figure}[hbt]  \vspace{2.5in}
  \includegraphics{fig84c.ps}
\end{figure}

\newpage 

FIGURE 2b

\begin{figure}[hbt]  \vspace{2.5in}
  \includegraphics{fig84logc.ps}
\end{figure}

\clearpage

TABLE 1

\begin{table}[h] \[ \hspace*{-3mm} \begin{array}{rclcl}
\Lambda X_1              & \rightleftharpoons &   X_2^\gamma 
&& \mbox{uncatalysed, with forward-rate coeff.} = \ep \\
\Lambda X_1 + X_2^\gamma & \rightleftharpoons & 2 X_2^\gamma
&& \mbox{autocatalysis, with forward-rate coeff.} = \alpha \\ 
\Lambda X_1 + X_2^\theta   & \rightleftharpoons & 
X_2^\gamma + X_2^\theta
&& \mbox{crosscatalysis, with forward-rate coeff.} = \chi  \\
\Lambda X_1 + X_2^\gamma + X_2^\theta   
& \rightleftharpoons & 2 X_2^\gamma + X_2^\theta
&& \mbox{enzymatic-catalysis, with forward-rate coeff.} =\zeta . 
\end{array} \hspace*{-3mm} \] \end{table}

\clearpage

TABLE 2

\[ \hspace*{-9mm} \begin{array}{rclccl}
\Lambda X_1 & \rightarrow & Y_n && \ep 
        &  \mbox{ uncatalysed formation of short chains} \\ 
Y_n + \Lambda X_1 & \rightarrow & Y_{n,m} && \ep 
        &  \mbox{ uncatalysed formation of long chains} \\  &&&&& \\ 
Y_n + \Lambda X_1 & \rightarrow & 2 Y_n && \alpha 
        & \mbox{ autocatalysis of short chains} \\ 
Y_{m,n} + Y_m + \Lambda X_1 & \rightarrow & 2 Y_{m,n} && \alpha 
        & \mbox{ autocatalysis of long chains} \\  &&&&& \\ 
Y_m + \Lambda X_1 & \rightarrow & Y_m + Y_n && \chi 
        & \mbox{ crosscatalysis of short by short} \\ 
Y_{m,n} + Y_k + \Lambda X_1 & \rightarrow & Y_{m,n} + Y_{k,l} && \chi 
        & \mbox{ crosscatalysis of long by long} \\
Y_{m,n} + \Lambda X_1 & \rightarrow & Y_{m,n} + Y_k && \chi 
        & \mbox{ crosscatalysis of short by long} \\
Y_m + Y_k + \Lambda X_1 & \rightarrow & Y_m + Y_{k,l} && \chi 
        & \mbox{ crosscatalysis of long by short} \\ &&&&& \\ 
Y_p + Y_n + \Lambda X_1 & \rightarrow & 2 Y_n + Y_p && \zeta 
        & \mbox{ ribozymic synthesis of short chains} \\ 
Y_{p,q} + Y_n + \Lambda X_1 & \rightarrow &  2 Y_n + Y_{p,q} && \zeta 
        & \mbox{ ribozymic synthesis of short chains} \\ 
Y_p + Y_{m,n} + \Lambda X_1 & \rightarrow & Y_n  +  Y_{m,n} + Y_p && \zeta 
        & \mbox{ ribozymic synthesis of short chains} \\ 
Y_{p,q} + Y_{m,n} + \Lambda X_1 & \rightarrow & Y_n +  Y_{m,n} + Y_{p,q} &&
\zeta 
        & \mbox{ ribozymic synthesis of short chains} \\ 
Y_p + Y_{n,m} + \Lambda X_1 & \rightarrow & Y_n + Y_{n,m} + Y_p && \zeta 
        & \mbox{ ribozymic synthesis of short chains} \\ 
Y_{p,q} + Y_{n,m} + \Lambda X_1 & \rightarrow & Y_n +  Y_{n,m}+Y_{p,q} &&
\zeta 
        & \mbox{ ribozymic synthesis of short chains} \\  &&&&& \\ 
Y_p + Y_{m,n} + Y_m + \Lambda X_1 & \rightarrow  & 2 Y_{m,n} + Y_p && \zeta
        & \mbox{ ribozymic synthesis of long chains} \\
Y_{p,q} + Y_{m,n} + Y_m + \Lambda X_1 & \rightarrow  & 2 Y_{m,n} + Y_{p,q}
&& \zeta 
        & \mbox{ ribozymic synthesis of long chains} \\
Y_p + Y_{m,n} + Y_n + \Lambda X_1 & \rightarrow  & 2 Y_{m,n} + Y_p && \zeta
        & \mbox{ ribozymic synthesis of long chains} \\
Y_{p,q} + Y_{m,n} + Y_n + \Lambda X_1 & \rightarrow  & 2 Y_{m,n} + Y_{p,q}
&& \zeta 
        & \mbox{ ribozymic synthesis of long chains} 
\end{array} \hspace*{-9mm} \] 

\clearpage

TABLE 3

\begin{table}[h] \[ \hspace*{-9mm} \begin{array}{rclccl}
\Lambda X_1 & \rightarrow & Y_n && \epS 
        &  \mbox{ uncatalysed polymerisation} \\ 
Y_n + \Lambda X_1 & \rightarrow & Y_{n,m} && \epL 
        &  \mbox{ uncatalysed growth of long from short chains} \\  
Y_n + \Lambda X_1 & \rightarrow & 2 Y_n && \alphaS 
        & \mbox{ autocatalysis of short chains} \\ 
Y_{m,n} + Y_m + \Lambda X_1 & \rightarrow & 2 Y_{m,n} && \alphaL 
        & \mbox{ autocatalysis of long chains} \\  
Y_{m,n} & \rightarrow & Y_m + Y_n  &&  \eta & \mbox{ hydrolysis } \\ 
                                                &&&&& \\ 
Y_m + \Lambda X_1 & \rightarrow & Y_m + Y_n && \chiS 
        & \mbox{ crosscatalysis of short by short chains} \\ 
Y_{m,n} + Y_k + \Lambda X_1 & \rightarrow & Y_{m,n} + Y_{k,l} && \chiL 
        & \mbox{ crosscatalysis of long by long chains} \\
Y_{m,n} + \Lambda X_1 & \rightarrow & Y_{m,n} + Y_k && \chiX 
        & \mbox{ crosscatalysis of short by long chains} \\
                                                &&&&& \\ 
Y_{p,q} + Y_n + \Lambda X_1 & \rightarrow &  2 Y_n + Y_{p,q} && \zetaSS 
        & \mbox{ ribozymic synthesis of short chains} \\ 
Y_{p,q} + Y_{m,n} + \Lambda X_1 & \rightarrow & Y_n +  Y_{m,n} +
Y_{p,q}&&\zetaSL
        & \mbox{ ribozymic synthesis of short chains} \\ 
Y_{p,q} + Y_{n,m} + \Lambda X_1 & \rightarrow & Y_n +  Y_{n,m}+Y_{p,q}
&&\zetaSL
        & \mbox{ ribozymic synthesis of short chains} \\  
Y_{p,q} + Y_{m,n} + Y_m + \Lambda X_1 & \rightarrow  & 2
Y_{m,n}+Y_{p,q}&&\zetaLL
        & \mbox{ ribozymic synthesis of long chains} \\
Y_{p,q} + Y_{m,n} + Y_n + \Lambda X_1 & \rightarrow  &
2Y_{m,n}+Y_{p,q}&&\zetaLL
        & \mbox{ ribozymic synthesis of long chains} 
\end{array} \hspace*{-9mm} \] \end{table}

\clearpage

TABLE 4

\begin{table}[h] \[ \hspace*{-3mm} \begin{array}{rclccl}
Y_m + Y_k + \Lambda X_1 & \rightarrow & Y_m + Y_{k,l} && \chi 
        & \mbox{ crosscatalysis of long by short chains} \\ 
Y_p + Y_n + \Lambda X_1 & \rightarrow & 2 Y_n + Y_p && \zeta 
        & \mbox{ ribozymic synthesis of short chains} \\ 
Y_p + Y_{m,n} + \Lambda X_1 & \rightarrow & Y_n  +  Y_{m,n} + Y_p && \zeta 
        & \mbox{ ribozymic synthesis of short chains} \\ 
Y_p + Y_{n,m} + \Lambda X_1 & \rightarrow & Y_n + Y_{n,m} + Y_p && \zeta 
        & \mbox{ ribozymic synthesis of short chains} \\ 
Y_p + Y_{m,n} + Y_m + \Lambda X_1 & \rightarrow  & 2 Y_{m,n} + Y_p && \zeta
        & \mbox{ ribozymic synthesis of long chains} \\
Y_p + Y_{m,n} + Y_n + \Lambda X_1 & \rightarrow  & 2 Y_{m,n} + Y_p && \zeta
        & \mbox{ ribozymic synthesis of long chains} 
\end{array} \hspace*{-3mm} \] \end{table}


\renewcommand{\baselinestretch}{1.0}
\small
\newpage


\newpage

\addcontentsline{toc}{section}{References}

\begin{thebibliography}{99}
%
\bibitem{eigen} Eigen, M. 
{\em Die Naturwissenschaften}, {\bf 1971}, 58, 465.
%
\bibitem{Gilbert} Gilbert, W. 
{\em Nature}, {\bf 1986}, 319, 618. 
%
\bibitem{orgel} Orgel, L.E. 
{\em Scientific American} Oct {\bf 1994}, 271, 77.
%
\bibitem{ch95} Coveney, P.V.; Highfield, R. {\em 
Frontiers of Complexity}; Faber \& Faber: London, 1995.
%
\bibitem{mss} Maynard Smith, J.; Szathm\'{a}ry, E. {\em The Major
Transitions in Evolution}; Oxford University Press, 1997.
%
\bibitem{joyce89} Joyce, G.F. {\em Nature}, {\bf 1989},338, 217.
%
\bibitem{joy-org} Joyce, G.F.; Orgel, L.E. {\em Prospects 
for Understanding the Origin of the RNA World}; in 
The RNA World; eds Gesteland, R.F.; Atkins, J.F.; 
Cold Spring Harbor Laboratory Press: New York,1993; pp 1-26. 
%
\bibitem{fleisch} Fleischaker, G.; Colonna, S.; Luisi, P.L. 
{\em Self-Production of Supramolecular Structures}. NATO ASI Series,
vol.~446; Kluwer: Dordrecht, 1994.
%
\bibitem{eigenpop} Eigen, M. {\em Steps Towards Life. A 
Perspective on Evolution}; Oxford University Press: Oxford, 1992.
%
\bibitem{ems88} Eigen, M.; McCaskill, J.; P Schuster, P.  
{\em J. Phys. Chem.}, {\bf 1988}, 92, 6881. 
%
\bibitem{eigmccsch} Eigen, M.; McCaskill, J.; Schuster, P.;  
{\em Adv. Chem. Phys.}, {\bf 1989}, 75, 149. 
%
\bibitem{eigenetal} Eigen, M.; Schuster, P. {\em The hypercycle. A 
principle of natural self-organisation}; Springer-Verlag: New York, 1979.
%
\bibitem{cn} Chacon, P.; Nu\~{n}o, J.C.  
{\em J. Biological Systems}, {\bf 1995}, 3, 351.
%
\bibitem{cnp} Chacon, P.; Nu\~{n}o, J.C.   
{\em Physica D}, {\bf 1995}, 81, 398.
%
\bibitem{ne2} Nu\~{n}o, J.C.; Andrade, M.A.; Montero, F. 
{\em Bull. Math. Biol.}, {\bf 1993},  55, 417.
%
\bibitem{ne1} Nu\~{n}o, J.C.; Andrade, M.A.; Moran, F.; Montero, F. 
{\em Bull. Math. Biol.}, {\bf 1993}, 55, 385. 
%
\bibitem{nmr} Nu\~{n}o, J.C.; Montero, F.; de la Rubia, F.J.
{\em J. Theor. Biol.}, {\bf 1993}, 165,  553.
%
\bibitem{nt} Nu\~{n}o, J.C.; Tarazona, P.
{\em Bull. Math. Biol.} {\bf 1994}, 56, 875. 
%
\bibitem{bd35} Becker, R.; D\"{o}ring, W. 
{\em Ann. Phys.}, {\bf 1935}, 24, 719.
%
\bibitem{pl76} Penrose, O.; Lebowitz, J.L.  In  {\em Studies in 
Statistical Mechanics VII: Fluctuation Phenomena};  ed Montroll, E.; 
Lebowitz, J.L.; North Holland: Amsterdam, 1976; pp.322-375. 
%
\bibitem{cw96} Coveney, P.V.; Wattis, J.A.D.   
{\em Proc. Roy. Soc. Lond. A}, {\bf 1996}, 452, 2079.
%
\bibitem{cw98} Coveney, P.V.;  Wattis, J.A.D.  
{\em J. Chem. Soc.: Faraday Transactions}, {\bf 1998}, 94, 233. 
%
\bibitem{wc97} Wattis, J.A.D.; Coveney, P.V. 
{\em J. Chem. Phys.}, {\bf 1997}, 106, 9122. 
%
\bibitem{dold} Dold, J.W. 
{\em Proc. Roy. Soc. Lond. A}, {\bf 1991}, 433, 521.
%
\bibitem{Smol} von~Smoluchowski, M. 
{\em Physik. Z.}, {\bf 1916}, 17,  557.
%
\bibitem{bcp86} Ball, J.M.; Carr, J.; Penrose, O. 
{\em Communs Math. Phys.}, {\bf 1986}, 104, 657.
%
\bibitem{Carr} Carr, J. {\em Applications of Centre Manifold
Theory}; Springer-Verlag: New York, 1981.
%
\bibitem{as72} Abramowitz, M. ; Stegun, I.A. {\em Handbook of
Mathematical Functions}; Dover: New York, 1972. 
%
\bibitem{Murray} Murray, J.D. {\em Mathematical Biology}. 
Biomathematics, Vol 19; Springer-Verlag: Berlin, 1989; Appendix 2. 
%
\bibitem{falini} Falini, G.; Albeck, S.; Weiner, S.; Addadi, L. 
{\em Science}, {\bf 1996}, 271, 67. 
%
\bibitem{kam} Kam, Z.; Shore, H.B.; Feher, G. 
{\em J. Mol. Biol.}, {\bf 1978}, 123, 539.
%
\bibitem{beckmann} Beckmann, W.; Otto, W.H. 
{\em Chem. Eng. Res. \& Des.}, {\bf 1996}, 74, 750.
%
\end{thebibliography}
\end{document}